\documentclass[12pt,preprint]{aastex}

\begin{document}

\title{Physical Conditions in the Narrow-Line Region of Markarian 3.
         I. Observational Results}

\author{Nicholas R. Collins\altaffilmark{1}, Steven B. Kraemer\altaffilmark{1},
D. Michael Crenshaw\altaffilmark{2}, Jose Ruiz\altaffilmark{1},
Rajesh Deo\altaffilmark{2}, and Frederick C. Bruhweiler\altaffilmark{1}}

\altaffiltext{1}{Institute for Astrophysics and Computational
Sciences, Catholic University of America; and NASA Goddard Space
Flight Center, Code 681, Greenbelt, MD 20771;
collins@stis.gsfc.nasa.gov,  stiskraemer@yancey.gsfc.nasa.gov, 
ruiz@yancey.gsfc.nasa.gov, fredb@iacs.gsfc.nasa.gov}
\altaffiltext{2}{Department of Physics
and Astronomy, Georgia State University, Astronomy Offices,
One Park Place South SE, Suite 700, Atlanta, GA  30303;
crenshaw@chara.gsu.edu, deo@chara.gsu.edu}

\begin{abstract}
We use Hubble Space Telescope/Space Telescope Imaging Spectrograph
(HST/STIS) longslit low-resolution spectroscopy from 1150~\AA~
to 10,300~\AA~ to study the physical conditions in the narrow-line
region (NLR) of the Seyfert~2 galaxy Markarian~3.
We find from the \ion{He}{2}~$\lambda$1640/$\lambda$4686 line
ratio and the Balmer decrement that the extinction within Markarian~3 
along the line-of-sight to the NLR
is best characterized by a Large Magellanic Cloud (LMC)
type extinction curve.
We observe an extinction gradient increasing from west to east
along the STIS slit (at position angle
71$\degr$ measured east from north)
in both line and continuum emission.  We infer from this gradient
that the host galaxy disk is tilted towards the observer
in the east: the line-of-sight to the eastern emission-line cone intersects
more dust in the plane of the galaxy than that to the western
cone.
From emission-line diagnostics we find that the NLR
gas is photoionized by the hidden active galactic nucleus
(AGN) continuum and
that its density decreases with increasing distance
from the center. 
We model the observed continuum as a combination of
reddened host galaxy light from an old stellar population,
reddened H$^{+}$ and He$^{+2}$ recombination continua, and
less reddened  scattered light from the central engine with
spectral index $\alpha$=1 (L$_{\nu}\propto\nu^{-\alpha}$).
The host galaxy to scattered-light ratio
is estimated to be 3:1 at 8125~\AA~ in 0$\farcs$1 $\times$ 1$\farcs$8
aperture.
Using a two-component power-law  model for the
ionizing continuum ($\alpha$=2 for 13.6~eV~$<$~E~$<$~0.2~keV
and $\alpha$=1 for 0.2~keV~$<$~E~$<$~50~keV) we find
that the covering factor (normalized for our observation
aperture) of the NLR gas is $>$0.7\%.  We estimate
that the amount of intrinsic non-ionizing UV continuum
scattered into our line-of-sight is 0.04\%.  This is
consistent with our estimate of the scattering fraction
for broad \ion{C}{4}~$\lambda\lambda$1548,1551 emission.
\end{abstract}


\section{Introduction}

Markarian 3 is classified as a Hubble-type SB0 galaxy \citep{ada77}
with a Seyfert~2 (Sy~2) active galactic nucleus (AGN) \citep{wee69}.
Its spectrum is characterized by narrow forbidden and permitted emission lines
and excess UV flux compared to normal galaxies \citep{wee73}.
\citet{tho80}, \citet{sch85}, \citet{mil90} and \citet{tra95} observed
broad permitted lines and AGN continuum emission
in its polarized spectrum.
These observations are part of the foundation
for the ``unified'' model for active galaxies
(Antonucci 1993; Urry \& Padovani 1995). In this model
the presence or absence of broad permitted-emission lines
and a bright, variable AGN continuum in the observed (unpolarized)
spectrum depends on
the viewing angle to the central engine.
The systemic velocity is 4050~km~s$^{-1}$ (z=0.0135) based on \ion{H}{1}~21~cm
emission \citep{tif88}. The distance to Markarian is 53~Mpc if
H$_{0}$=75~km~s$^{-1}$~Mpc$^{-1}$.
At this distance, 1$\arcsec$ corresponds to 257~pc.

Markarian~3 has been well studied from X-rays
to radio wavelengths.  Its NLR has a reverse-``S'' shape
greater than 2$\arcsec$ in extent \citep{cap95}.
In a study of near-UV to far-IR absorption lines,
\citet{gon01} found evidence of an old stellar population in the
nucleus of Markarian~3.
A 2$\arcsec$-long radio jet detected in 5-GHz MERLIN observations
\citep{kuk93} has a similar
shape and orientation to the NLR
morphology determined by HST/Faint Object Camera (FOC)
narrow-band imaging observations
centered on the $[$\ion{O}{2}$]\lambda~$3727,
$[$\ion{O}{3}$]\lambda~$5007, H$\alpha$, and H$\gamma$
emission-lines  \citep{cap95}.  Its hard X-ray
spectrum can be modeled as Compton reflection from molecular
gas in a torus that obscures the central source of ionizing
radiation (Georgantopoulos 1999; Turner et al 1997).  However,
\citet{geo99} found a factor of two variation in the X-ray
source flux between 4~keV and 20~keV over a 200-day period in
RXTE observations.  They inferred that a fraction of the
intrinsic X-ray continuum penetrates the putative obscuring
torus.   A comparison of ASCA, GINGA, and BBXRT data revealed that
the extended soft X-ray flux (E$<$4~keV), which covers the
same $\sim$12$\arcsec$ range as the extended narrow-line-region
(ENLR) \citep{sak00}, has shown no such variation over
$\sim$13 years \citep{iwa94}.

\citet{rui01} studied the kinematics of the Markarian~3 NLR using
$[$\ion{O}{3}$]$~$\lambda$5007 measurements from a STIS/CCD slitless
spectrum and from the same
longslit data that is presented in this
paper.  We use the results of their kinematic analysis in this paper.
The extracted spectra from several locations along the longslit
show evidence of two kinematic components: one is redshifted and
the other is blueshifted in the rest-frame of Markarian~3.  These
components are interpreted as line-of-sight observations of radially
outflowing gas on opposite sides of a NLR bi-cone.
Their best-fit kinematic model for a NLR with an angular extent
of 2$\arcsec$ and inner and outer opening half-angles of
15$\degr$ and 25$\degr$, respectively, is
radial acceleration from the nucleus to a turnover distance (measured
along the bi-cone axis) at
$\sim$80~pc followed by deceleration to the systemic velocity at
$\sim$260~pc.
The maximum space
velocity in the model is 1750~km/s.
In this model the bi-cone is tilted
towards the observer in the east and
away from the observer in the west by 5 degrees.
The position angle of the the bi-cone is 70$\degr$ east of
north \citep{sch00}.
We note that the host galaxy disk major-axis position angle
is 28$\degr$ and the inclination of the disk is
33$\degr$ \citep{sch00}.  We determine the direction of the
host-galaxy tilt in this paper.

In this paper we present an analysis of longslit spectra obtained
in each of the HST/STIS low-resolution modes: G140L, G230L, G430L,
and G750L.  These spectra show fifty emission-lines from
$\sim$21 ions.  The brightest emission-lines have
an angular extent of $\sim$3$\farcs$4 or a projected linear
extent of 870~pc.
In \S2 we describe the observations and data reduction.
In \S3 we describe the emission-line measurement procedure and discuss
the extinction corrections.  In \S4 we present
temperature, density and ionization  diagnostics.
In \S5 we show the observed continuum and describe its
constituents.  We also estimate the covering factor
of the NLR gas, and the scattering fraction for the
continuum produced by the hidden AGN.  Our physical
picture of the NLR, including the relative orientation
of the host galaxy disk, is presented in \S6.  We
summarize our results in \S7.  In the forthcoming paper II,
we will use these observations and the photoionization model code
CLOUDY by \citet{fer88}
to constrain the sizes, densities, and spatial distribution
of gas clouds in the NLR.

\section{Observations and Data Reduction}

\begin{figure}
\figurenum{1}
\begin{minipage}{5in}
\plotone{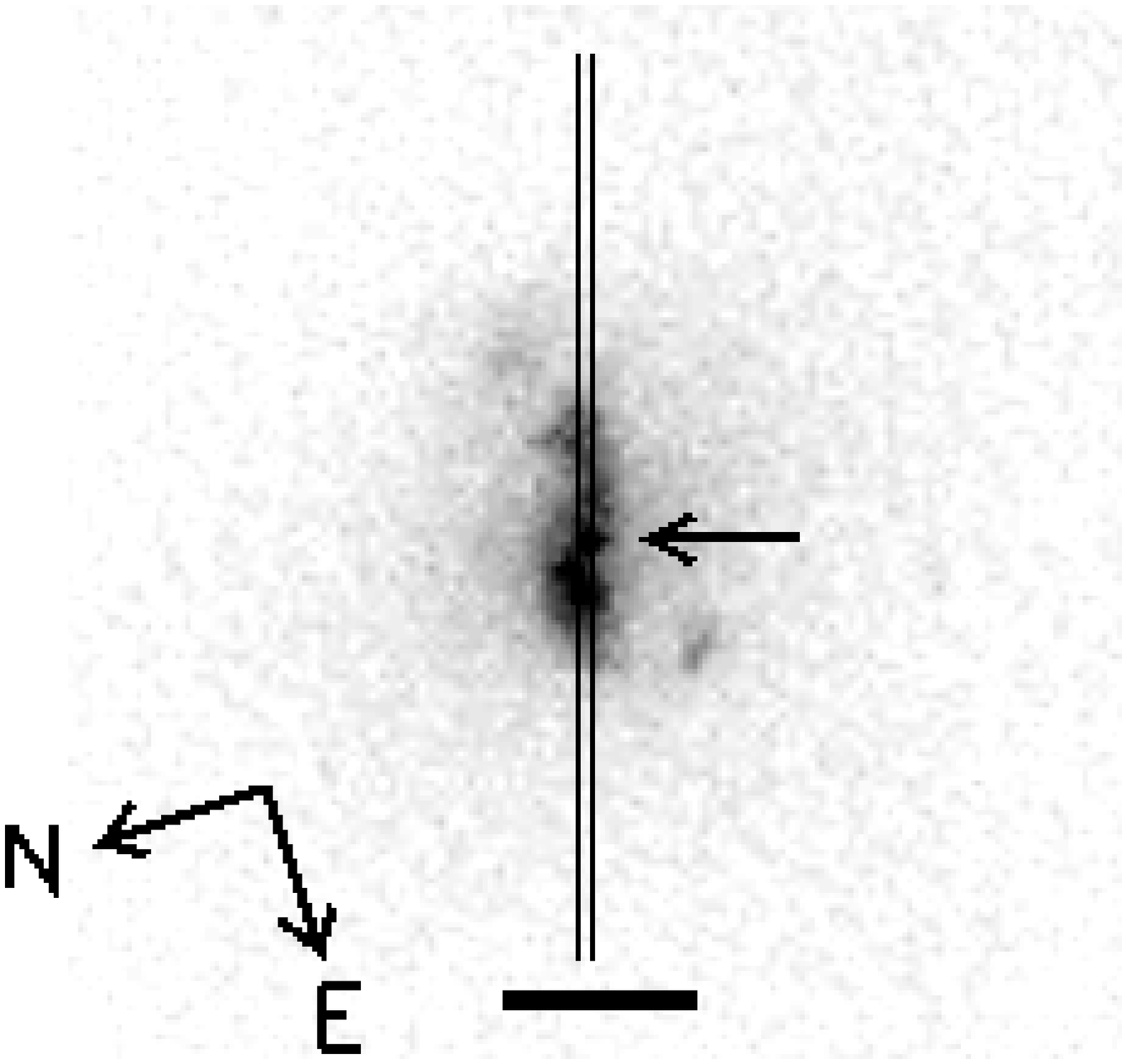}       
 \caption{STIS target acquisition image obtained
through the F28X50LP filter with 52$\arcsec$x0$\farcs$1 slit
superposed.  The horizontal bar drawn below the NLR is 1$\arcsec$ or 257~pc
in length. The slit is oriented 71$\degr$ east of north.  East is
towards the bottom of the figure and North is towards the left.
The arrow indicates the NLR kinematic center derived by \citet{rui01}
and the position of the putative hidden nucleus identified by \citet{kis02}.
} \label{slide1c}
\end{minipage}
\end{figure}

HST/STIS longslit spectra were obtained on 22 August 2000 in all
four low-resolution modes, G140L, G230L, G430L, and G750L
providing wavelength coverage from 1150~\AA~ to 10,300~\AA.
The observations are listed in Table~1.
The 52$\arcsec\times$0$\farcs$1 slit was used for all spectral modes.
This aperture is superposed on the target
acquisition image in Figure~\ref{slide1c}.  The slit was oriented
to position-angle 71$\degr$ east of north to match the elongated
reverse-``S'' shaped feature in the NLR.  East is towards the bottom
of this figure, and North is towards the left, as indicated by the arrows.
The target acquisition image was obtained  through the STIS long-pass F28X50LP
filter, which spans red to near-IR wavelengths.  The bright knots
in the image are primarily emission from lines that fall within the
wavelength range of the F28X50LP filter.  The brightest lines 
in the visual and near-IR are
$[$\ion{O}{1}$]~\lambda\lambda$6300, 6363,
$[$\ion{N}{2}$]~\lambda\lambda$6548, 6584,
H$\alpha$~$\lambda$6563, 
$[$\ion{S}{2}$]~\lambda\lambda$6716, 6731,
and $[$\ion{S}{3}$]~\lambda\lambda$9069, 9532.

The spectra were processed with the CALSTIS IDL software developed by
\citet{lin98} for the STIS Instrument Definition Team.  This software
is used to perform bias and dark subtraction, hot pixel correction,
flat-field division, and for the CCD images, cosmic-ray rejection.
Hot pixels were identified in STIS dark images and corrected by
interpolating in the dispersion direction.  Cosmic-ray artifacts were
removed in the CCD data by comparing the multiple image readouts
obtained within each spectral mode.  The multiple readouts obtained in
the G140L, G430L, and G750L modes were combined to increase their
signal-to-noise ratios.  The wavelength scales were corrected for
zero-point shifts using hole-in-the-mirror (HITM) lamp calibration
exposures made after each science exposure.  The two G140L spectra
have a relative wavelength offset of 0.15~\AA~, or 0.25~pixel.  Both
spectra were shifted to match a central wavelength of 1418.821~\AA.
No such shifts were required for the mode G430L and mode G750L input spectra.

The spectra were processed in ``extended-source'' mode.  In this mode,
each column of the output spectral image corresponds to a unique
wavelength and each row corresponds to a unique spatial position.
This rectification was performed using a bi-linear interpolation
algorithm \citep{lin98}.  The units of the output spectral image are
ergs~s$^{-1}$~cm$^{-2}$~\AA$^{-1}$~cross-dispersion-pixel$^{-1}$.  In
addition to this rectification, all spectral images were transformed
to the same spatial plate scale (1~pixel = 0$\farcs$05071) and
alignment as that of spectral mode G430L.  This simplifies comparison
of the same NLR clouds observed in different spectral modes.  The
co-alignment among modes in the spatial direction
was determined by cross-correlating the slit
fiducial marks that are evident in the respective wavelength
calibration spectra.

Finally, a background correction was made for each of the spectral
images.  For modes G230L, G430L, and G750L, the background subtraction
algorithm follows that described by \citet{lin98} for point-source
extractions.  Briefly, two eleven-pixel wide strips are selected a
suitable distance away (typically more than 200 pixels, or 10$\arcsec$)
from the center of the NLR spectrum.  Each strip is averaged in the
spatial direction, then the
two average background spectra are in turn averaged together.  The
resultant background spectrum was smoothed three times: once with a
median filter, and again with two different mean filters.  This
smoothed background spectrum was subtracted from each line of the
spectral image.

The background subtraction for mode G140L required more care due to
contamination from geo-coronal Ly$\alpha$~$\lambda$1216 and
\ion{O}{1}~$\lambda$1303 emission  .  The pixels
in the averaged, smoothed background spectrum that contain the two
geo-coronal lines were replaced with pixels from the averaged,
unsmoothed background spectrum.  This method minimizes subtraction
residuals at the positions of these lines.

We examined the impact of fringing  on the CCD/G750L mode spectrum.
Fringing in the STIS/CCD G750L mode spectra can reach peak-to-peak
amplitude variations of greater than 20\% at $\lambda>$~9000~\AA~
\citep{pro02,mal03}.  The fringing is caused by interference among
reflected waves in the CCD substrate.  We investigated the fringe
amplitude of our data using a tungsten lamp flat-field \citep{pla00}
and a model flat-field \citep{mal03}.  We concluded that the
uncertainty in the data in the wavelength region affected by fringing
($\lambda >$ 7500~\AA) is dominated by Poisson noise rather than
fringing.  No correction for fringing was deemed necessary.

\begin{figure}[t]
\figurenum{2}
\begin{minipage}{6.25in}
\plotone{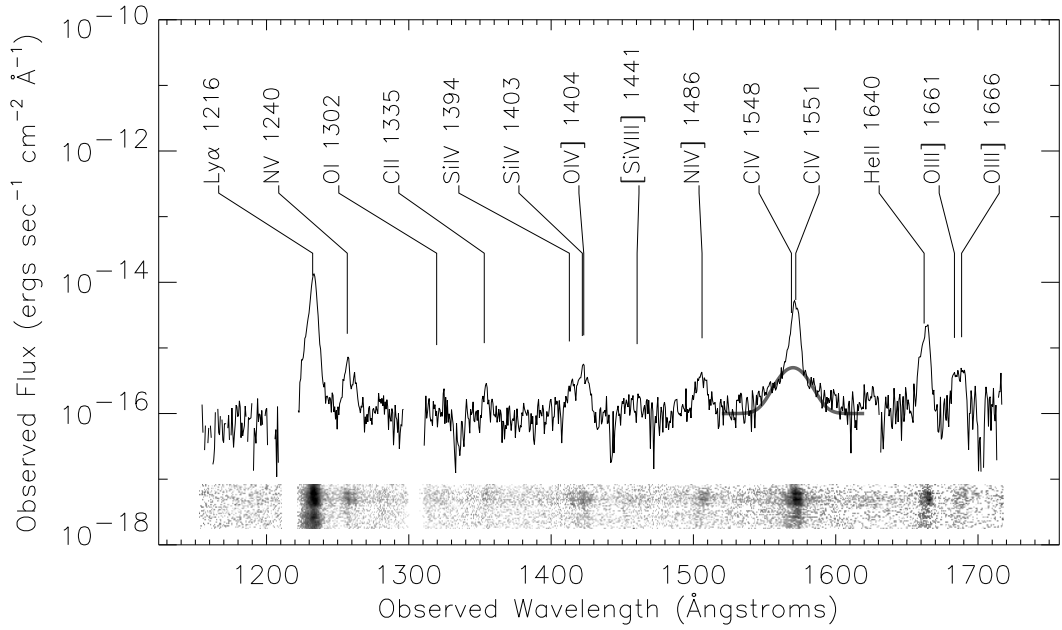}      
\caption{1$\farcs$8 extraction from G140L spectral image. The heavy
gray line at \ion{C}{4}~$\lambda\lambda$1548,1551 shows the
fit to the broad-line flux scattered into our line-of-sight
that is discussed in \S5.2. The 1$\farcs$8 spectral image extraction 
is displayed at the bottom of the panel.}
\label{slide6}
\end{minipage}
\end{figure}

\begin{figure}[b]
\figurenum{3}
\begin{minipage}{6.25in}
\plotone{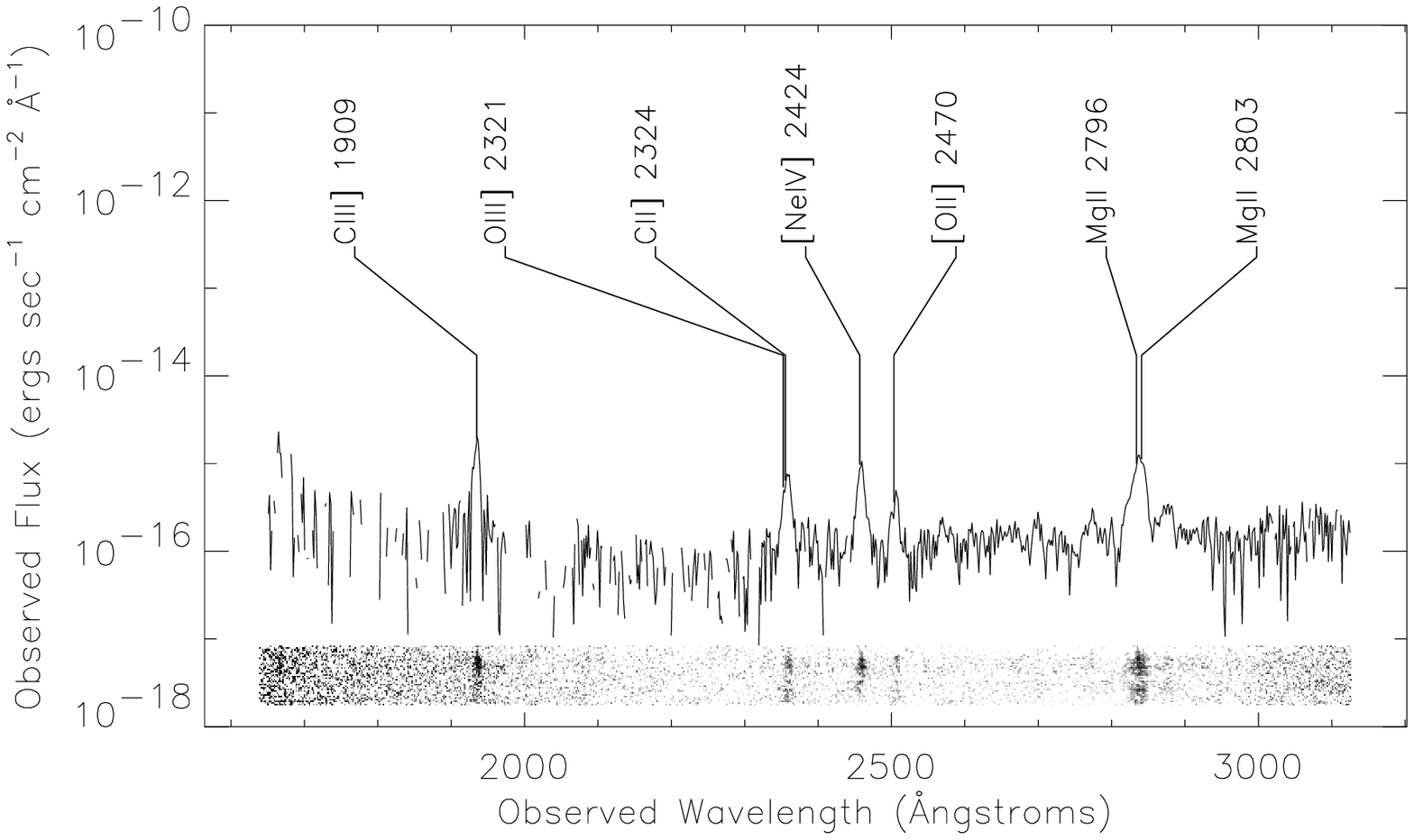}      
\caption{1$\farcs$8 extraction from G230L spectral image.} \label{slide7}
\end{minipage}
\end{figure}

\begin{figure}[t]
\figurenum{4}
\begin{minipage}{6.25in}
\plotone{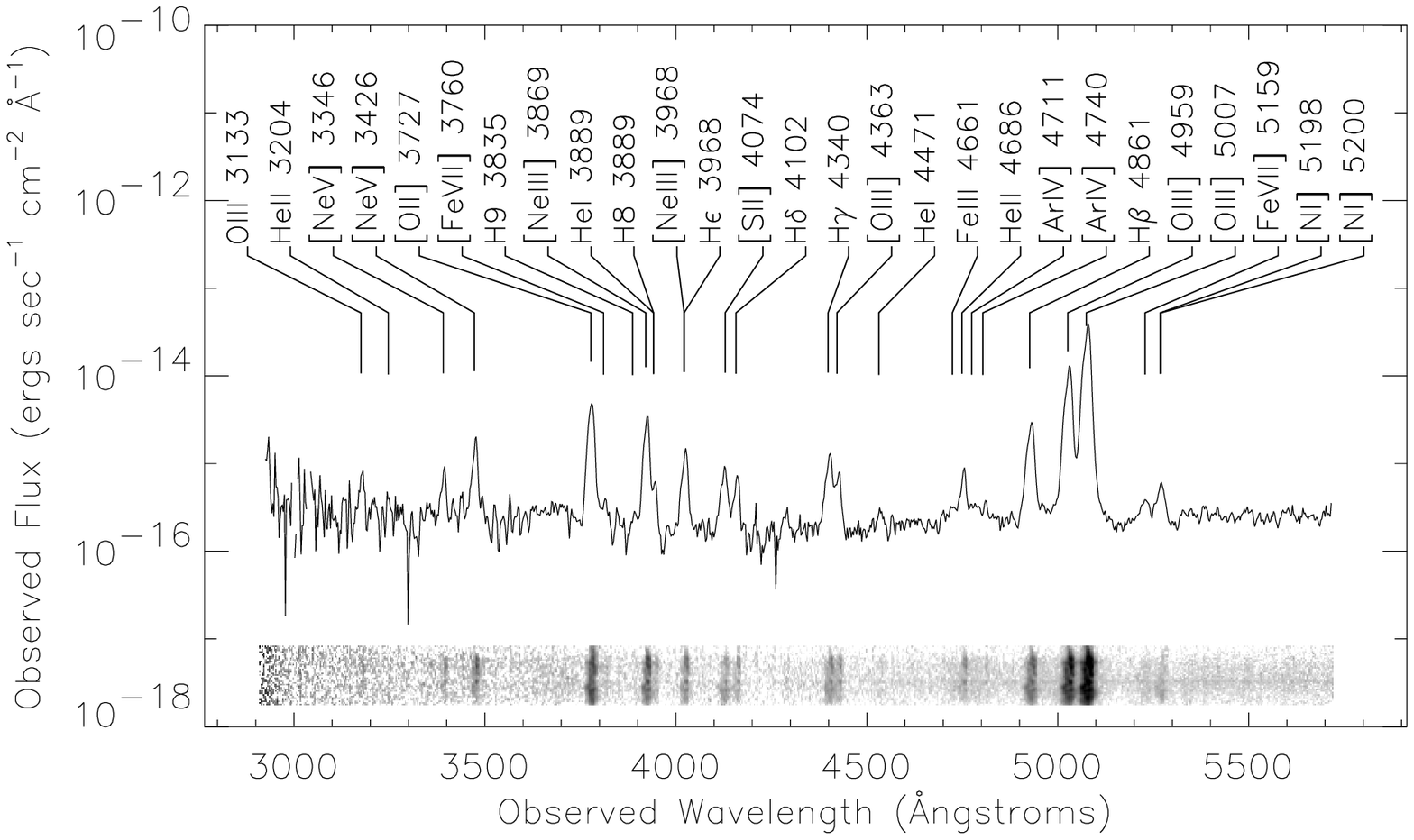}      
\caption{1$\farcs$8 extraction from G430L spectral image.} \label{slide8}
\end{minipage}
\end{figure}

\begin{figure}[b]
\figurenum{5}
\begin{minipage}{6.25in}
\plotone{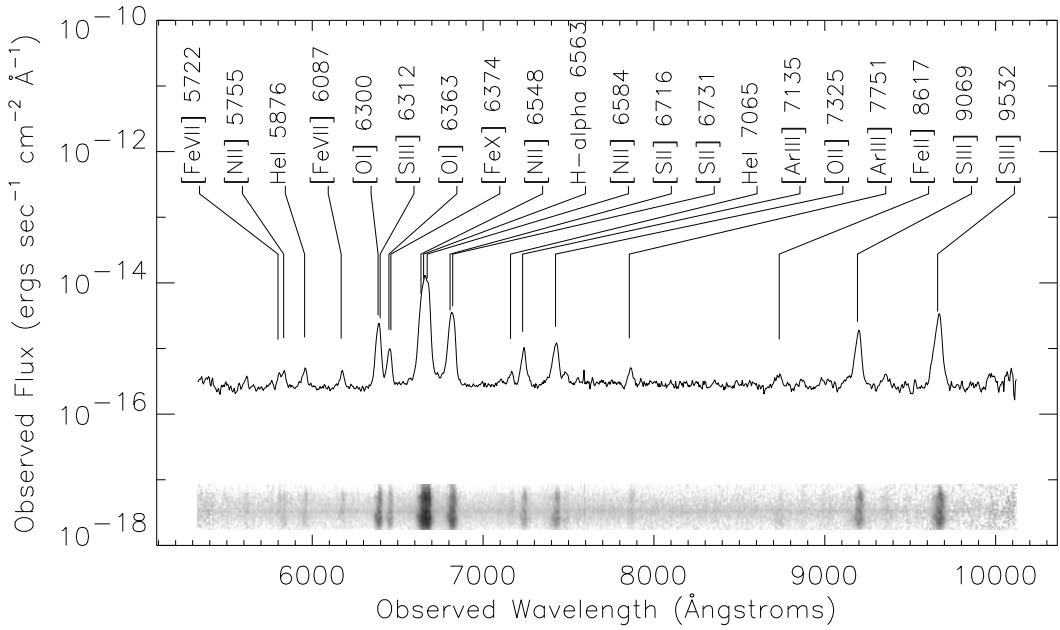}      
\caption{1$\farcs$8 extraction from G750L spectral image.} \label{slide9}
\end{minipage}
\end{figure}

We display 1$\farcs$8 sub-image extractions from the co-aligned,
background-subtracted spectral images for G140L,
G230L, G430L and G750L in Figures
\ref{slide6}, \ref{slide7}, \ref{slide8}, and \ref{slide9}, respectively.
The overlaid 1-D spectrum is the sum
of the spectral image in the cross-dispersion direction.
The spatial orientation of each spectral image is West at the top of
the figure and East at the bottom of the figure, consistent with
Figure~\ref{slide1c}. We detected over 60 emission lines from ionic
species that require ionization potentials up to 100~eV.


\section{Emission-Line Measurements}

\subsection{Integrated Fluxes}

We made seven discrete measurements of each emission line
in the spatial direction.  Individual spectra were extracted by
summing array rows in selected bins in the spatial direction.  The bin
sizes were selected to extract spectral information for individual
clouds and to achieve a reasonable signal-to-noise ratio for the the
\ion{He}{2}~$\lambda$1640/$\lambda$4686 measurement which
is used to determine the extinction along the line-of-sight to the
NLR (see \S3.2).
The bin sizes range from 0.2-0.3 arcseconds, or 52-77 pc, using the
adopted distance scale described in the Introduction.   The full
spatial range along the slit spanned by the defined bins was
also constrained by the spatial extent of the two \ion{He}{2}
emission lines.  Several emission-lines have spatial extents
greater than the analyzed 1$\farcs$8 range, but the extinction
correction outside this range cannot be determined from the low
signal-to-noise \ion{He}{2} emission.

\begin{figure}
\figurenum{6}
\begin{minipage}{6.25in}
\plotone{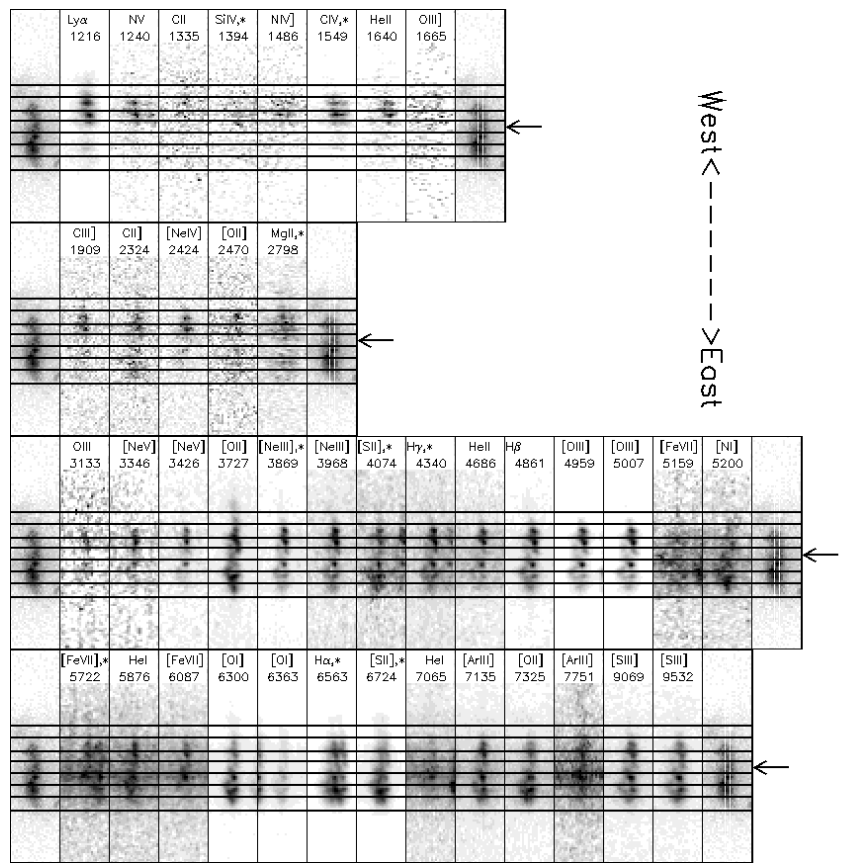}      
\caption{Measurement bin diagram shown for 50 emission lines.
The spatial orientation is west at the top and east at the bottom.
The spatial range from the bottom of the eastern-most bin 
to the top of the western-most bin is 1$\farcs$8. 
The spectral range shown for each line is 
15~\AA~ for mode G140L (top row), 
40~\AA~ for mode G230L (second row), 
68~\AA~ for mode G430L (third row), 
and 123~\AA~ for mode G750L (bottom row)
The target acquisition image (see Figure~1) is duplicated at the ends of
each row for
reference. At the right end of each row, the 52$\arcsec\times$0$\farcs$1 slit
boundaries are drawn on the target acquisition image.
The asterisks next to some of the line labels indicate that
more than one emission line is present in the box.
The arrow indicates the bin defined as 0$\farcs$0 in 
this paper.  
} \label{slide11}
\end{minipage}
\end{figure}

The measurement bins are shown in Figure~\ref{slide11}.
The bins were selected
by visually comparing the spatial profile of each emission-line with
that of the target acquisition direct image.
The arrow on
each row of displayed emission lines indicates bin 0$\farcs$0. 
It contains the
kinematic zero-point in the \citet{rui01} velocity map.  This 
coincides with the position of the hidden-nucleus derived by \citet{kis02}
from UV-imaging polarimetry data.

The line flux measurement procedure began by removing 
all lines except the line of interest from the one-dimensional 
spectral extraction.  We executed this removal by interpolating 
between continuum values adjacent to each unwanted line.  
We then performed a linear fit to the remaining continuum. 
We subtracted this fit from the spectrum, then fitted a Gaussian 
function to the line of interest.  The uncertainties in the line fluxes were
estimated by making three such measurements for each line,
each time interactively selecting different continuum regions.
The wavelength range of the continuum regions varied between the
full spectral range of the spectral mode of interest
(G140L: 611~\AA, G230L: 1616~\AA, G430L: 2800~\AA, and
G750L: 5030~\AA) and one-third of the the full range.
The reported line flux is the average value of the three
measurements.
The line flux uncertainty is the standard deviation
of the three measurements added in quadrature with the
uncertainties from Poisson statistics.  This uncertainty
is dominated by the standard deviation of the three line flux
measurements.

\begin{figure}
\figurenum{7}
\begin{minipage}{6.25in}
\plotone{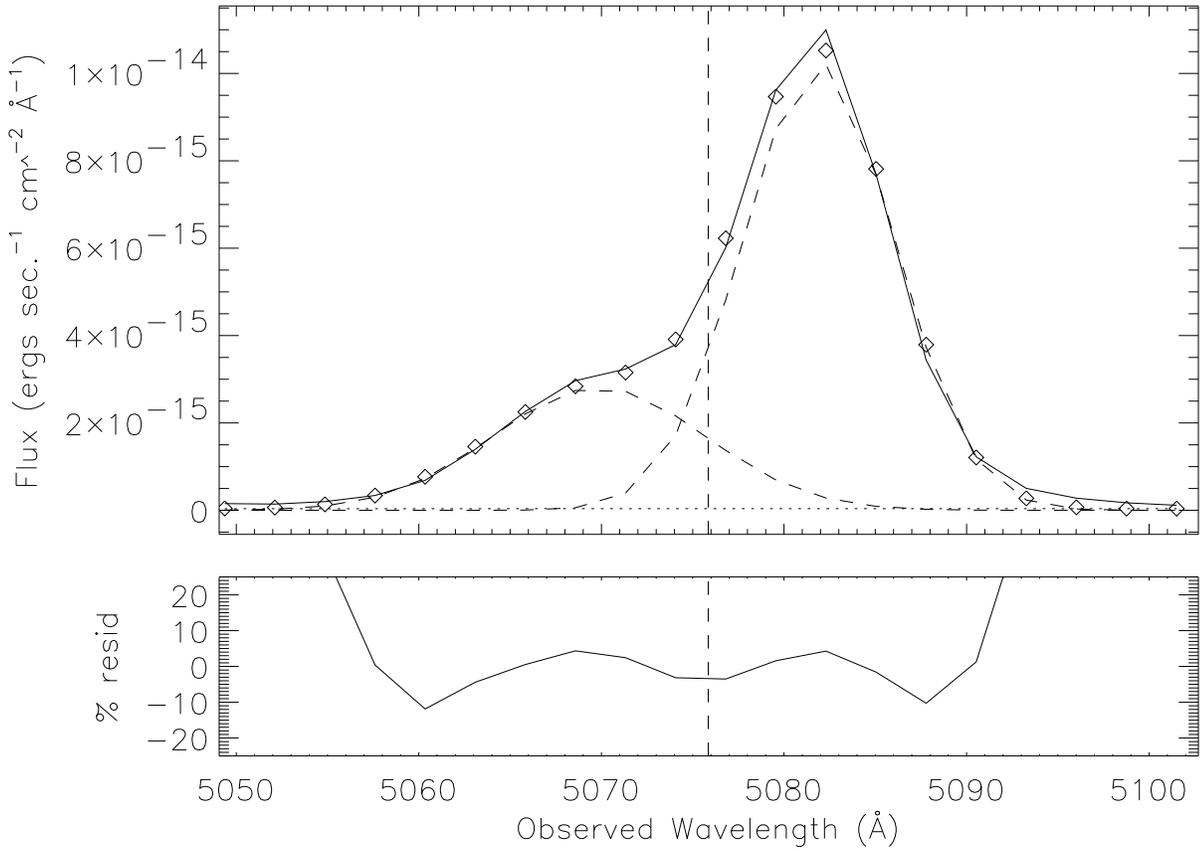}      
\caption{Two component Gaussian fit to the $[$\ion{O}{3}$]$~$\lambda$5007
line in the central bin (position = 0$\farcs$0).
The dashed vertical line indicates the
wavelength of $[$\ion{O}{3}$]$~$\lambda$5007 at the systemic velocity
of Markarian~3.  The dotted line near the bottom of the emission-line shows
the fit to the continuum.  The dashed lines show the fits to the two
emission-line components.  The diamonds show the sum of the emission-line
and continuum fits.
} \label{o3fit}
\end{minipage}
\end{figure}

In any measurement bin where two kinematic components and/or two
different lines are blended together, we used the
$[$\ion{O}{3}$]$~$\lambda$5007 solution for that bin as a template.
There were no compelling incidences where three or more components
were required to fit the measurements using the selected bins. We
assumed that the kinematics of the $[$\ion{O}{3}$]$~$\lambda$5007 gas
matches that of the line of interest.  The
$[$\ion{O}{3}$]$~$\lambda$5007 measurements with two components were
fitted with six free parameters, or three for each of the two
components: the peak flux, the Gaussian standard deviation, and
the central wavelength. The Gaussian standard deviations were
converted to velocity dispersions. The fitted central wavelengths
were used with the rest wavelength to compute the radial velocities
of the components. The dispersions were converted to Gaussian
standard deviations for the line(s) of interest.  These standard
deviations were corrected to match the observed line profiles by
adding in quadrature the appropriate spectral resolution term from
Table~1.  The radial velocities were used to derive the wavelength
offsets for the line(s) of interest. This procedure leaves only
the peak fluxes of the components as free parameters in the
Gaussian fits for the line(s) of interest.
The lines deblended in this manner were
          $[$\ion{Ne}{3}$]$~$\lambda$3869 from  H8~$\lambda$3889,
          $[$\ion{S}{2}$]$~$\lambda$4074 from H$\delta$~$\lambda$4102,
          H$\gamma$~$\lambda$4340 from $[$\ion{O}{3}$]$~$\lambda$4363,
          $[$\ion{Fe}{7}$]$~$\lambda$5159 from
          $[$\ion{N}{1}$]$~$\lambda$5200,
          $[$\ion{Fe}{7}$]$~$\lambda$5722 from
          $[$\ion{N}{2}$]$~$\lambda$5755,
          $[$\ion{O}{1}$]$~$\lambda$6300 from
          $[$\ion{O}{1}$]$~$\lambda$6363,
     and  \ion{He}{1}~$\lambda$7065 from
          $[$\ion{Ar}{3}$]$~$\lambda$7136.
For the
H$\alpha$~$\lambda$6563+$[$\ion{N}{2}$]$~$\lambda\lambda$6548,6584
blend we fixed the
$[$\ion{N}{2}$]$~$\lambda$6584/$\lambda$6548 flux ratio to 3:1  in
addition to the aforementioned constraints.  The 3:1 ratio applies
to a $\sim$10,000~K nebulae with density less than 10$^{5}$~cm$^{-3}$.
This is the critical density required to collisionally de-excite the
upper level ($^{1}$D$_{2}$)  responsible for the
$[$\ion{N}{2}$]$~$\lambda\lambda$6548,6584 doublet emission.
We used the same procedure to deblend the \ion{C}{4}~$\lambda\lambda$1548,1551
doublet and the \ion{Mg}{2}~$\lambda\lambda$2796,2803 doublet.  For
both of these doublets the short-to-long wavelength line-flux ratio
constraints are 2:1 for optically thin lines.
The spectral resolution is too low to separate
the individual components of the
remaining blended
lines: \ion{Si}{4}~$\lambda$1398 multiplet with the
\ion{O}{4}$]$~$\lambda$1402 multiplet,
      and $[$\ion{S}{2}$]$~$\lambda$6716 with
          $[$\ion{S}{2}$]$~$\lambda$6731.
Figure~\ref{o3fit} shows the fit for the two-component
$[$\ion{O}{3}$]$~$\lambda$5007 line in the central bin.
We list flux measurements relative to H$\beta$~$\lambda$4861
for forty-six emission lines at seven spatial positions in Table~\ref{tab:2}.  The
1$\sigma$ uncertainties in the observed flux ratios are listed in
parentheses below each measurement.


\subsection{Extinction Correction}

We corrected the observed flux measurements for both
Galactic extinction along the line-of-sight to Markarian~3, and
intrinsic extinction caused by dust within Markarian~3.
The Galactic color-excess of
E(B-V)~=~0.19 was obtained from the dust map of \citet{sch98}. We
used the \citet{sav79} extinction curve to correct for Galactic
extinction.

We used the \ion{He}{2}~$\lambda$1640/$\lambda$4686 line ratio
\citep{sea78} to determine the extinction within Markarian~3 along
the line-of-sight.  This ratio has a longer wavelength baseline
than the Balmer decrement
(H$\alpha$~$\lambda$6563/H$\beta$~$\lambda$4861) and therefore
gives a better estimate of the reddening.  It is also insensitive
to collisional excitation.  The ratio can vary from 5.6 for a low
density (n$_{e}$ $\rightarrow$ 0~cm$^{-3}$) 5,000~K plasma to 7.6
for a high density (n$_{e}$$\sim$10$^{6}$~cm$^{-3}$) 20,000~K
plasma \citep{sea78}.  We used a ratio of 7.2. This is appropriate
for a 10,000~K nebula with Case~B recombination and n$_{e}$ $<$
10$^{4}$~cm$^{-3}$ \citep{sea78}.
We calculated the color excess
for each measured NLR component using the 
following screen geometry relation  
\begin{equation}
E(B-V) = 2.5 \times log_{10}(7.2 \times (f_{4686}/f_{1640})) /(X_{1640} - X_{4686})
\end{equation}
where f$_{4686}$ and f$_{1640}$ are the Galactic extinction corrected
\ion{He}{2} line fluxes
and X$_{1640}$ and  X$_{4686}$ are the normalized extinction values
(X$_{\lambda}$ $\equiv$ E($\lambda$ - V) / E(B-V))
at the wavelengths
of the \ion{He}{2} lines.  
The 1$\sigma$ uncertainties in the 
color excess values are on the order of the range of color excess 
values calculated using the theoretical upper and lower limits of the 
\ion{He}{2}~$\lambda$1640/$\lambda$4686 line ratio described above. 

To determine the best extinction curve for the NLR of
Markarian~3, we calculated E(B-V) using several different curves 
from the literature, applied the different color excess values 
to the H$\alpha$~$\lambda$6563 and H$\beta$~$\lambda$4861 
flux measurements, and compared the resultant Balmer decrement 
values with theoretical values.  The Balmer decrement 
can vary from 2.74 for a high density (n$_{e}$ = 10$^{6}$~cm$^{-3}$) 
20,000~K gas to 3.05 for a low density (n$_{e}$ $\rightarrow$ 0) 
5,000~K gas assuming Case B recombination \citep{ost89}. 
We also required physically 
plausible values (typically $\leq$36 [Osterbrock 1989]) 
for the extinction corrected Ly$\alpha$~$\lambda$1216 to 
H$\beta$~$\lambda$4861 flux ratios.
We used following extinction curves: Galactic \citep{sav79} ,
$\theta^{2}$~Ori~B \citep{boh81}, the starburst curves of
\citet{cal94} and \citet{cal97}, SMC-bar \citep{wit99}, SMC
\citep{hut82}, average LMC \citep{nan81}, average LMC
\citep{koo81},  and separate curves for within and outside of
30~Doradus in the LMC \citep{fit85}. 

The $\theta^{2}$~Ori~B, Galactic, starburst,  
SMC-bar curves yielded values  too low to be
physically plausible.  The Balmer decrement weighted 
mean values and their uncertainties are, respectively, 
0.44$\pm$0.08, 2.33$\pm$0.09, 2.30$\pm$0.13, 2.36$\pm$0.10, and 
2.02$\pm$0.11. The SMC curve \citep{hut82}  yielded 
a Balmer decrement weighted mean value of 2.84$\pm$0.06, 
but the extinction corrected Ly$\alpha$~$\lambda$1216 values
were higher than the theoretical upper limit by factors of up to 2.8.
The \citet{nan81} and \citet{fit85} LMC curves 
(within and without 30~Doradus) produced
weighted mean Balmer decrement values  closer to the theoretical limits:  
2.53$\pm$0.08, 2.52$\pm$0.08 (30~Dor), and 2.35$\pm$0.08 (non-30~Dor), 
respectively.  We found that the LMC curve of \citet{koo81}  
gave a weighted mean value of the Balmer decrement (2.59$\pm$0.07) 
closest to the theoretical values and yielded physically plausible 
values for the Ly$\alpha$~$\lambda$1216 to H$\beta$~$\lambda$4861 
flux ratios.  We used this curve to de-redden the
observed fluxes using a screen geometry in this paper.
The extinction corrected line fluxes relative to H$\beta$~$\lambda$4861
are listed in Table~3.  The 1$\sigma$ line-ratio uncertainties 
include a component due to the uncertainty in the extinction correction.

The simple grain model of \citet{mat77} for standard Galactic
extinction curves attributes the continuous extinction to a
size-distribution of various grains, and the 2200~\AA~ bump
to smaller (mainly graphite) grains. The empirically
derived \citet{koo81} LMC curve exhibits a less pronounced
2200~\AA~ bump and increases more steeply with
decreasing wavelength for $\lambda <$ 2200~\AA.
The steeper rise of the \citet{koo81} LMC curve is
attributed to an overall smaller distribution of dust grains
than are required to produce the Galactic, starburst, and
SMC-bar extinction curves described in this section.  The agreement of the
\ion{He}{2}~$\lambda$1640/$\lambda$4686 ratios and the Balmer decrement
values with their theoretical
values after applying the  \citet{koo81} LMC curve suggests
that the dust grain size distribution in the observed area of
the Markarian~3 NLR is smaller than the standard
Galactic distribution. \citet{pit00} note that many
active galaxies have dust extinction characterized by the
absence of the 2200~\AA~ feature.
The reddening curve is generally similar to that of
the narrow-line Sy~1 Arakelian~564 \citep{cre02}, but not as steep as
that of the SMC \citep{hut82}.

\begin{figure}
\figurenum{8}
\begin{minipage}{6.25in}
\plotone{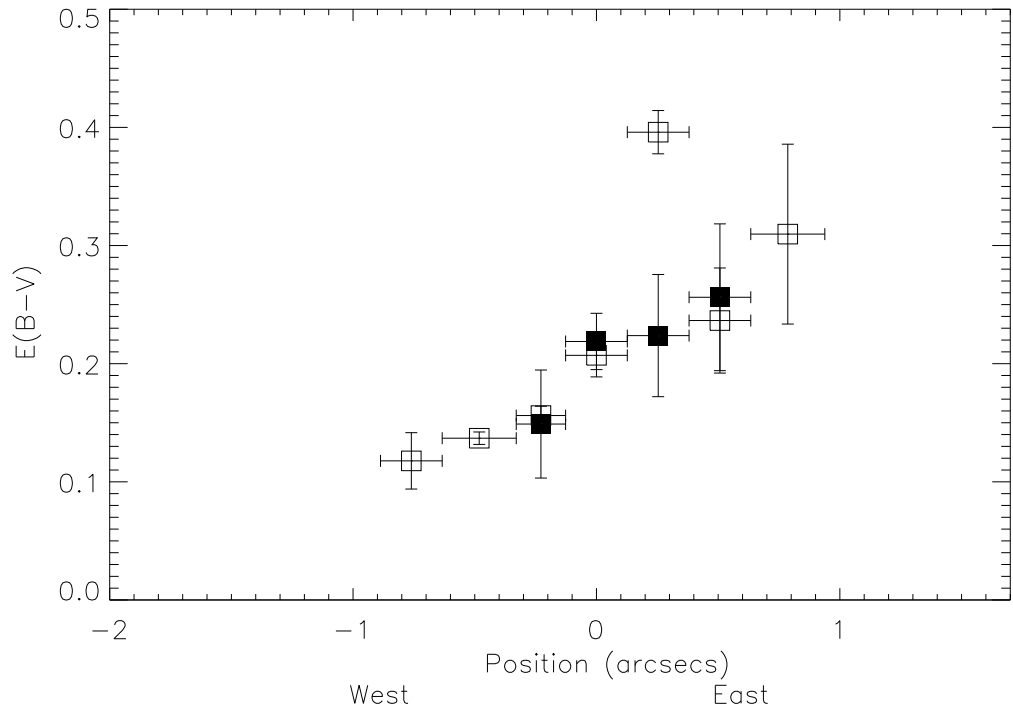}      
\caption{E(B-V) color excess as a function of  angular separation
from the nucleus along the slit.  Vertical bars show the uncertainty in E(B-V),
and horizontal bars show the angular size of the measured bin.
Open squares represent
redshifted components and filled squares represent blueshifted
components.  There is a clear extinction gradient increasing
from west to east in the NLR. The zero-point of the position
axis for Figures~\ref{fig1}-\ref{fig12} is the kinematic center
determined by \citet{rui01}.} \label{fig1}
\end{minipage}
\end{figure}

The color excess is shown as a function of position in the NLR in
Figure~\ref{fig1}.  There is a striking gradient in extinction
increasing from west to east.  This result might also have been
inferred from inspection of the spectral images.  Compare the
spatial morphology of UV emission lines in Figures~\ref{slide6} and
\ref{slide7} with that of  optical lines in Figures~\ref{slide8} and
\ref{slide9}.  Both redshifted and blueshifted
points east of the nucleus are reddened, suggesting that the source of
extinction may be an intervening dust lane or screen between the
observer and the NLR along the line-of-sight.
The extinction gradient
suggests that there is more dust in the screen along the line-of-sight to
the eastern side of the NLR bi-cone 
than that of the western side, as observed along the STIS slit.
\citet{elm99} noted that dust
in an inclined galactic plane can produce a similar extinction
gradient in which the side tilted towards the observer is more
reddened than the side that is tilted away (see further discussion on
the host galaxy orientation in \S6).
We note that the magnitude of the extinction gradient is steeper  
when derived using an extinction curve that increases slowly 
with shorter wavelength in the UV, 
such as the $\theta^{2}$~Ori~B \citep{boh81} 
curve.  The \citet{hut82} SMC curve, which increases faster 
with decreasing wavelength in the UV, produces a shallower 
extinction gradient than that derived with the LMC \citet{koo81} 
curve. 

\citet{kra00} found no clear trend over their measured region 
($\pm$3$\arcsec$, or $\sim\pm$200~pc for z=0.0033 and 
H$_{0}$=75~km~s$^{-1}$~Mpc$^{-1}$) 
of the NLR in NGC~4151 (Sy~1.5), although they note that the
extinction may increase from the north-east to the southwest over the
inner $\pm$1$\arcsec$ ($\sim\pm$60pc).  
They concluded that the lack of
uniformity in extinction over the entire NLR suggests that the reddening is
associated with the emission-line knots, rather than an external
screen.  Note that the orientation of the NGC~4151 NLR is such
that all of the north-east side is redshifted and all of the
south-west side is blueshifted \citep{nel00}.  This is not surprising since 
the NLR bi-cone axis is tilted further toward our line-of-sight 
(35$\degr$ out of the plane of the sky [Evans et al. 1993]) 
than that in Markarian~3.  The host NGC~4151 galaxy is closer to face-on 
orientation than Markarian~3 with an inclination of 21$\degr$ such that 
the eastern side is tilted towards 
the observer \citep{kin00}. 

In NGC~1068 (Sy~2), \citet{kc00b} found in general that measurements
of blueshifted clouds exhibited greater extinction than redshifted
clouds.  They inferred that both red and blueshifted components were
observed through a foreground dust-screen, and that additional dust
may be associated with the individual blueshifted clouds. The
orientation of the NGC~1068 NLR is similar to that of Markarian~3,
where blue- and redshifted clouds are observed along the same
line-of-sight and the bi-cone axis is tilted towards the 
observer in the north-east by 5$\degr$ \citep{cre00}. 
The angular extent measured along the  the NLR bi-cone axis 
was $\pm$5$\arcsec$, or spatial extent $\pm$360~pc 
(for z=0.0038 and H$_{0}$=75~km~s$^{-1}$~Mpc$^{-1}$).
The host galaxy disk is tilted towards the observer 
in the south by 28$\degr$ \citep{kin00}.
The covering factors of the blueshifted
clouds are expected to be small so that those clouds do not eclipse
(and further redden) the redshifted clouds; therefore, the dust
responsible for the observed extinction gradient in the Markarian~3
NLR is likely a screen outside the NLR.  

We note that most of the emission line profiles in the dual-velocity-component
measurement bins (+0$\farcs$50, +0$\farcs$25, 0$\farcs$00, and
-0$\farcs$23) in the Markarian~3 STIS spectra show an asymmetry:
the observed flux of the blueshifted velocity component is less than
the flux of the redshifted component (e.g. Figure~\ref{o3fit}).
This asymmetry is observed
for resonance, semi-forbidden, and forbidden lines.  We do not
see evidence  in Figure~\ref{fig1} that the extinction is commensurately
greater for the blueshifted components compared to the redshifted
components.  The extinction corrected redshifted/blueshifted flux ratios are
also typically greater than unity.  The redshifted components may
be brighter simply because they have higher emissivity (i.e., more gas)
than the blueshifted components.

We  estimated the hydrogen column density along the
line-of-sight to the narrow-line region clouds using our derived
color excess values and the relation of \citet{shu85}:
N$_{HI}$=5.2$\times$10$^{21}$~cm$^{-2} \times $ E(B-V).
We found that the column density varies from
0.35$\times$10$^{21}$~cm$^{-2}$ to 1.8~$\times$10$^{21}$~cm$^{-2}$
corresponding to our range in calculated E(B-V) of 0.07 to 0.36.
These values are reasonable for an S0 galaxy viewed at
$i$=33 \citep{tuf04}, the inclination
of Markarian~3 \citep{sch00}.


\section{Emission-Line Diagnostics}

\subsection{Ionization Diagnostics}

\begin{figure}
\figurenum{9}
\begin{minipage}{6.25in}
\plotone{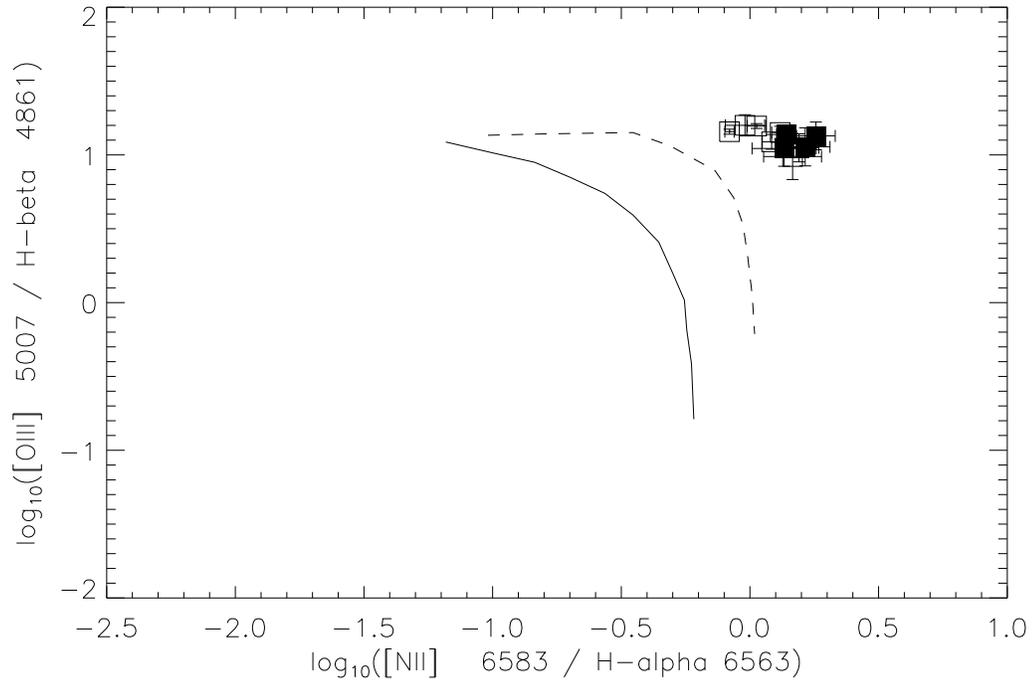}      
\caption{Diagnostic ratio plot to test the nature of the ionizing
continuum.  The solid line separates starbursts from AGN.  The
dashed line is the power-law photoionization model for solar
abundance of \citet{fer83}.
} \label{fig37}
\end{minipage}
\end{figure}

We used emission-line ratios to constrain the ionization mechanism
in the NLR. \citet{vei87} identified line ratios to distinguish
between starburst and AGN continua.
They used lines that are close in
wavelength to diminish the effects of reddening. We plot the logarithm
of the ratio $[$\ion{N}{2}$]~\lambda$6584/H$\alpha$~$\lambda$6563 against
that of $[$\ion{O}{3}$]$~$\lambda$5007/H$\beta$~$\lambda$4861 in
Figure~\ref{fig37}. The solid line
in each figure separates the starburst parameter space from that
for AGN \citep{vei87}.  The dashed line represents the power-law
photoionization model with solar abundance of \citet{fer83}.  The
dimensionless ionization parameter varies from 10$^{-4}$ in the lower
right to 10$^{-1.5}$ in the upper left.  The ionization parameter
is defined as U~=~Q(H)/4$\pi$$r^{2}n_{e}c$, where Q(H)
is the number of ionizing photons emitted by the central source
per second, $r$ is the distance between the source and the cloud,
$n_{e}$ is the electron density at the inner face of the cloud,
and $c$ is the speed of light.  If the density is low enough
that collisional de-excitation is negligible ($<$10$^{3}$~cm$^{-3}$
for most of the  observed emission-lines), emission-line
ratios will depend
primarily on U \citep{fer83}.  For example, the
$[$\ion{O}{3}$]$~$\lambda$5007/H$\beta$~$\lambda$4861
and the
$[$\ion{O}{3}$]$~$\lambda$5007/$[$\ion{O}{2}$]$~$\lambda$3727
ratios increase
as U increases.

The observations occupy a small region of parameter
space  that is approximately 0.2~dex higher
than the solar-abundance power-law photoionization model
of \citet{fer83}.   That the observations occupy a small
locus of parameter space indicates that the same
ionization mechanism, AGN photoionization,
applies for all the gas measured along the slit.
This is consistent with the results of \citet{kra86} and
of \citet{sak00} for Markarian~3.
The difference between the model and the data in the logarithmic ratio
corresponds to a factor of (0.2~dex=10$^{0.2}$=) 1.6.
This may be an indication that the N/O abundance ratio is
somewhat higher than solar (for solar abundances see
\citet{lam78}).  Such an offset is typical of Sy~2 NLR 
(see Figure 12.1 in \citet{ost89}). An enhanced N/O ratio 
may be an indication of a greater than solar nitrogen 
abundance combined with a 
less than solar oxygen abundance.  Oxygen may be depleted 
onto grains in a dusty nebula.  We will investigate 
this possibility in Paper II.

We also examined the line ratios
log($[$\ion{O}{1}$]~\lambda$6300/H$\alpha$~$\lambda$6563)
versus log($[$\ion{O}{3}$]$~$\lambda$5007/H$\beta$~$\lambda$4861).
This  indicator occupies a small region of its
AGN photoionization
parameter space similar to that of the
$[$\ion{N}{2}$]~\lambda$6584/H$\alpha$~$\lambda$6563 diagnostic. It
does not show the same offset from the \citet{fer83}
photoionization model however, suggesting that the abundance
for oxygen is close to solar.  This is also true
of the plots in \citet{ost89}.

\subsection{The Ionizing Continuum}

The \ion{He}{2}~$\lambda$4686/H$\beta$~$\lambda$4861 ratio is sensitive to
the shape of the ionizing continuum \citep{kra85,kra94} if the
NLR clouds are ionization-bounded.
The line ratio is lower with a steep spectrum than with a
flat spectrum because there are less He$^{+}$ ionizing photons
in the former case.  We assumed that neutral hydrogen is the
only absorber of photons with energies between 13.6~eV and 54.4~eV,
and that He$^{+}$ is the only absorber of photons with energies
greater than 54.4~eV.  For a power-law spectrum with spectral index $\alpha$
the ratio of the number of hydrogen ionizing photons, Q(H$^{0}$), to the
number of He$^{+}$ ionizing photons, Q(He$^{+}$), is
\begin{equation}
R = \frac{Q(H^{0})}{Q(He^{+})} = \frac{N(He)}{N(H)}
\frac{I(H\beta)}{I(4686)} \frac{j(4686)}{j(H\beta)}
\thickapprox \left( \frac{1}{4} \right)^{\alpha} - 1
\end{equation}
where N(He)/N(H) has the assumed solar value of $\sim$0.1,
I(H$\beta$)/I(4686) is the extinction-corrected measured line ratio,
and j(4686)/j(H$\beta$) is the ratio of volume emissivities
obtained from \citet{ost89}.
We summed the line fluxes measured in all bins for
\ion{He}{2}~$\lambda$4686 and H$\beta$ and found a ratio
I(H$\beta$)/I(4686) of 4.57. This yields
$\alpha$=1.36.  This agrees with the value of $\alpha$=1.4 we found for
NGC~1068 (Sy~2) \citep{kc00b},
NGC~5548 (Sy~1.5) \citep{cre99},
NGC~3783 (Sy~1) \citep{kcg01}, and
NGC~4151 (Sy~1.5) \citep{kc00a} in the spectral range 13.6~eV~$< E \lesssim$ 1~keV.
However, if there is a preponderance of matter-bounded
clouds, the intrinsic continuum could be steeper.

\subsection{Extinction-Corrected Line Ratios vs. Angular Separation from the
Kinematic Center}

\begin{figure}
\figurenum{10}
\begin{minipage}{6.25in}
\plotone{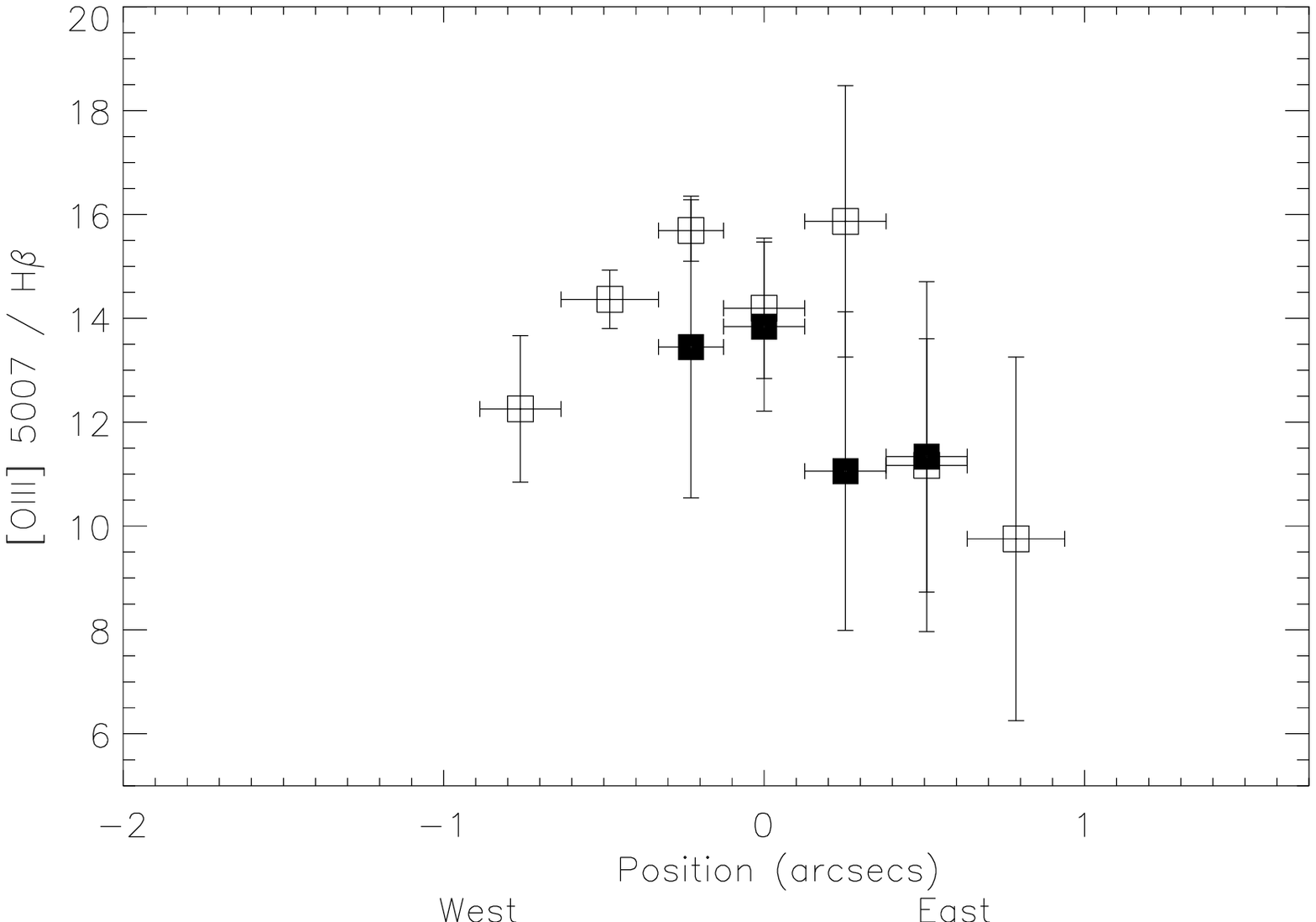}      
\caption{The ionization parameter sensitive line ratio
$[$\ion{O}{3}$]$~$\lambda$5007/H$\beta$~$\lambda$4861 vs.
 angular separation
from the nucleus along the slit.  Open squares represent redshifted components
and filled squares represent blueshifted components.
Generally, the ratio decreases with distance from the nucleus for
both redshifted and blueshifted components.
} \label{fig2}
\end{minipage}
\end{figure}

\begin{figure}[t]
\figurenum{11}
\begin{minipage}{6.25in}
\plotone{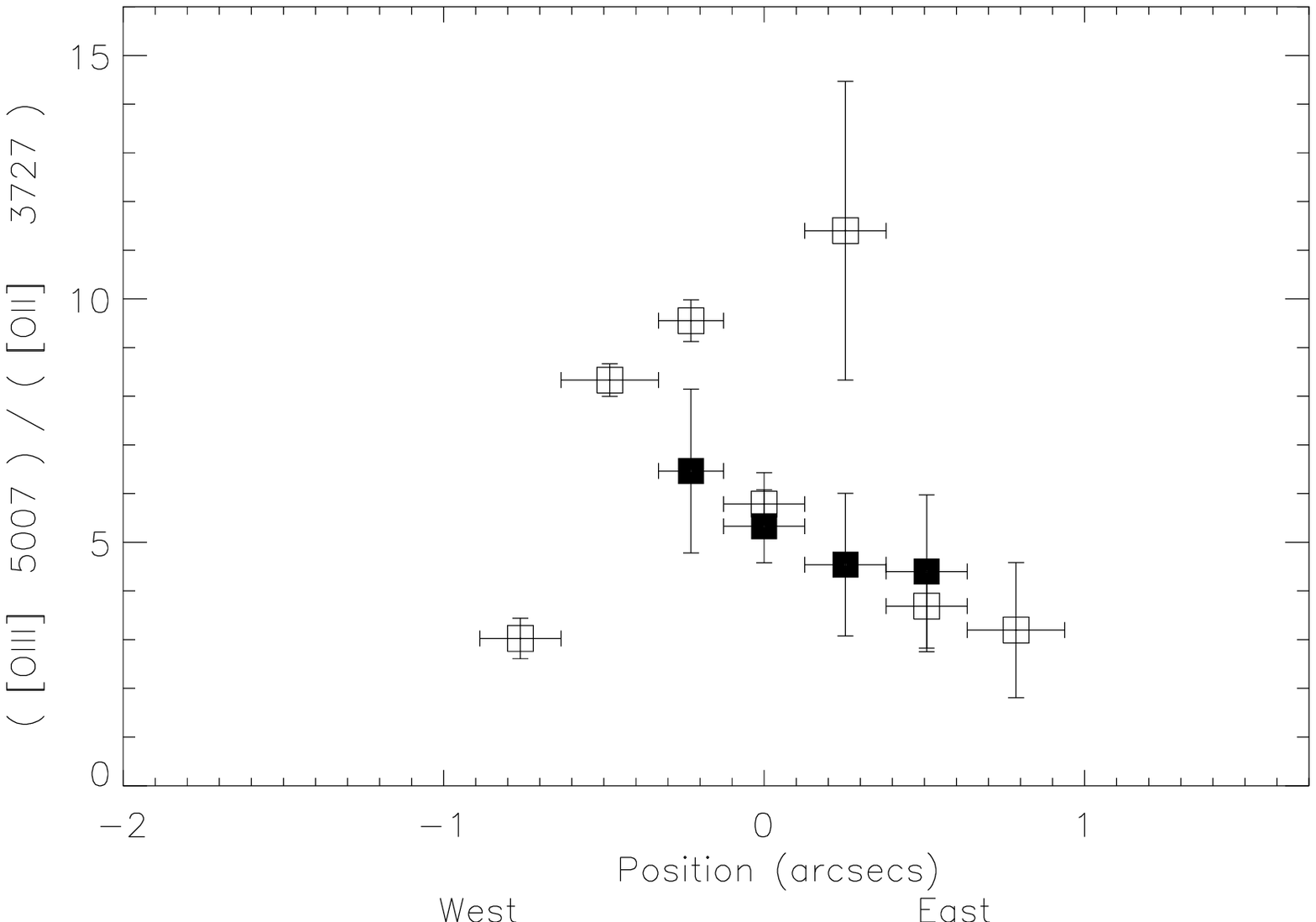}      
\caption{The ionization parameter sensitive line ratio
$[$\ion{O}{3}$]$~$\lambda$5007/$[$\ion{O}{2}$]$~$\lambda$3727~ vs. angular
separation
from the nucleus along the slit.  Open squares represent redshifted components
and filled squares represent blueshifted components.
The ratio decreases with distance from the nucleus for
both redshifted and blueshifted components.
} \label{fig4}
\end{minipage}
\end{figure}

\begin{figure}
\figurenum{12}
\begin{minipage}{6.25in}
\plotone{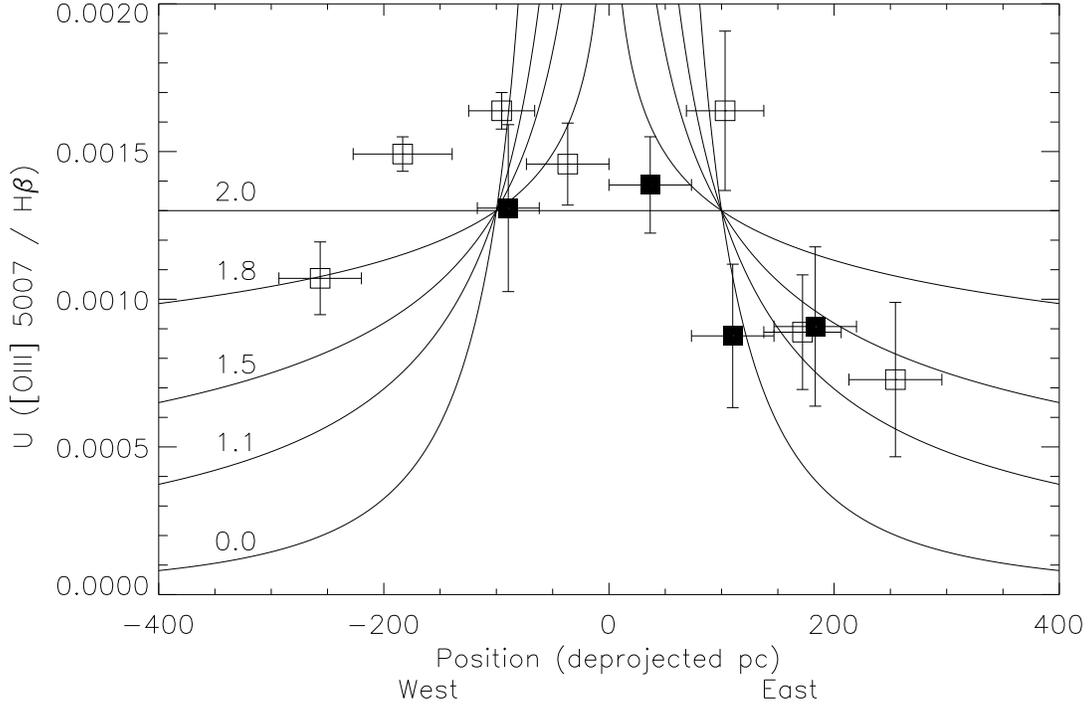}      
\caption{The ionization parameter sensitive line ratio
$[$\ion{O}{3}$]$~$\lambda$5007/H$\beta$~$\lambda$4861 vs.
radial distance from the nucleus towards the east and west 
along the slit.  Open squares represent 
redshifted components and filled squares represent 
blueshifted components. 
The curves represent theoretical ratios for different 
radial dependencies on electron density. 
The label above 
each curve refers to the exponential dependence 
on radius (for example, the curve labeled 2.0 represents 
a line ratio for which n$_{e}~\propto~$(1/r)$^{2.0}$). 
All curves are normalized to a ratio of 0.0013 at a radial 
distance of 100~pc west. 
} \label{uval_work_fig5}
\end{minipage}
\end{figure}

\begin{figure}[t]
\figurenum{13}
\begin{minipage}{6.25in}
\plotone{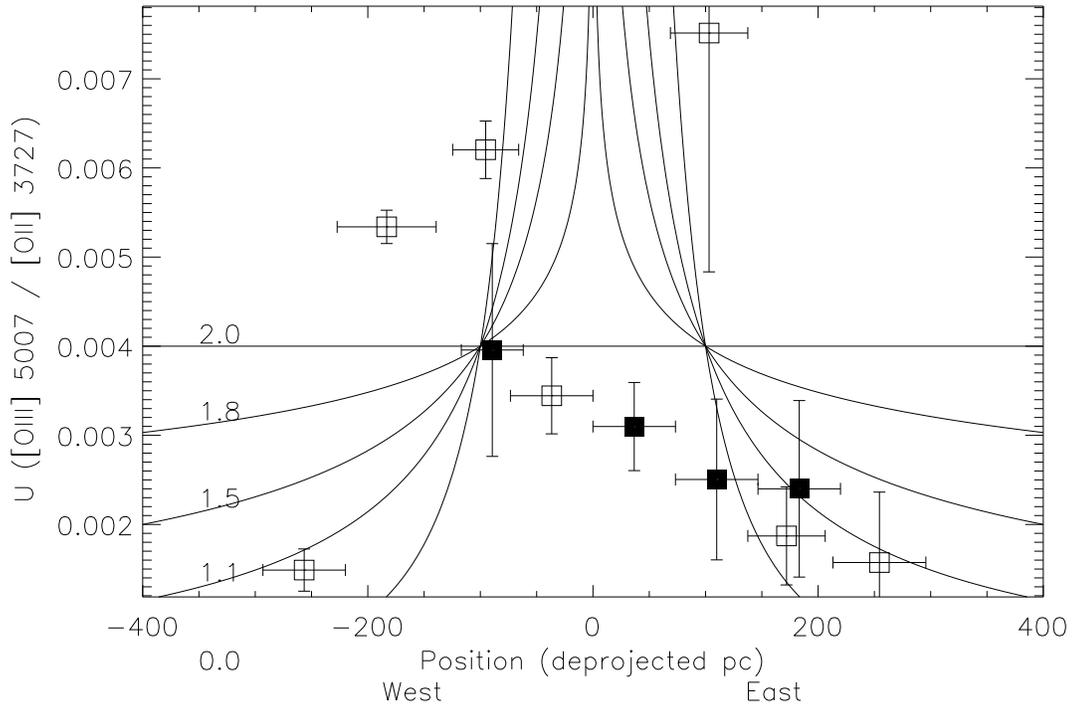}      
\caption{The ionization parameter sensitive line ratio
$[$\ion{O}{3}$]$~$\lambda$5007/$[$\ion{O}{2}$]$~$\lambda$3727~ vs. angular
radial distance from the nucleus towards the east and west 
along the slit. Symbols and curves are the same as those 
described in Figure~\ref{uval_work_fig5}. 
All curves are normalized to a ratio of 0.004 at a radial 
distance of 100~pc west. 
} \label{uval_work_fig6}
\end{minipage}
\end{figure}

The ionization-parameter sensitive line ratios can
be used to probe the electron density as a function
of position within the NLR. If $n_{e}$ is constant throughout the
NLR, the ionization parameter and its diagnostic line ratios
decrease as r$^{-2}$ in the NLR (see the definition of the ionization 
parameter, U, in \S4.1).  If $n_{e}$ varies with radius, 
the radial dependence of U (and its diagnostic line ratios) 
will be modified since U$\propto$(1/r)$^{2}\times$(1/$n_{e}$).
Figure~\ref{fig2} shows the extinction corrected flux ratio
$[$\ion{O}{3}$]$~$\lambda$5007/H$\beta$~$\lambda$4861 vs. projected
position in the NLR along the STIS longslit aperture.
A similar plot for the ratio
$[$\ion{O}{3}$]$~$\lambda$5007/$[$\ion{O}{2}$]$~$\lambda$3727
is shown in Figure~\ref{fig4}.
Using the bi-cone geometry of \citet{rui01} (inner and outer opening 
angles of 15$\degr$ and 25$\degr$, respectively, and 
tilted out of the plane of the sky by 5$\degr$ 
towards the observer in the east) and of \citet{sch96} 
(P.A.=70$\degr$)
we can determine 
the radial distances of the NLR clouds in Markarian~3.  
We use the solar abundance photoionization model 
of \citet{fer83} to convert the line ratios to photoionization 
parameter values.  The electron density is a constant 
10$^{3}$~cm$^{-3}$ for this model. 
The ionization parameter values derived from the two different 
line ratios are shown as functions of  radial distance  
in Figure~\ref{uval_work_fig5} and Figure~\ref{uval_work_fig6}. 
We show  curves on the figures representing U corresponding 
to different radial distributions of n$_{e}$.  
For the curve labeled ``2.0'', n$_{e}\propto$(1/r)$^{2.0}$
(and U is constant with radius), while 
n$_{e}$ is constant with radius (and U$\propto$(1/r)$^{2}$ 
for the curve labeled ``0.0''.  
Note that in the former case, the curve is flat because the radial 
dependence of the electron density cancels the explicit radial 
dependence of the ionization parameter. 
The values of U  at large radius in both Figures~\ref{uval_work_fig5} 
and \ref{uval_work_fig6} lie between the two plotted curves 
of n$_{e}\propto$(1/r)$^{2.0}$ and of n$_{e}$=constant.  This indicates 
that the electron density in the NLR decreases with radius from 
the nucleus.  The uncertainty in the ionization parameter is 
large, but the overall trends in both figures suggest that 
the decrease is slower than (1/r)$^{2}$.
A patchy gas distribution along the slit in the NLR might 
be responsible for deviations from the  curves. 
\citet{nel00} and
\citet{kai00} observed similar, although smoother, 
trends in STIS observations of NGC~4151  
and concluded that the electron density decreases
slower than r$^{-2}$ in its NLR.  

However, collisional de-excitation of the O$^{+}$ ions could
contribute to the higher line ratio close to the nucleus.
The critical densities of the upper-levels responsible for the
$[$\ion{O}{2}$]$~$\lambda$3727 doublet are
1.6$\times$10$^{4}$~cm$^{-3}$ for $^{2}$D$_{3/2}$ and
3.1$\times$10$^{3}$~cm$^{-3}$ for $^{2}$D$_{3/2}$ \citep{ost89}.
These are, respectively, factors of 44 and 226  less than the
critical density (7.0$\times$10$^{5}$~cm$^{-3}$)
for the upper level ($^{1}$D$_{2}$) responsible for the
$[$\ion{O}{3}$]$~$\lambda$5007 line \citep{ost89}.
If the electron density near the center of the NLR
were greater than the critical densities for the
$^{2}$D$_{3/2}$ and $^{2}$D$_{3/2}$ levels of O$^{+}$
and less than the critical density of the $^{1}$D$_{2}$
of O$^{+2}$, then the flux of the
$[$\ion{O}{2}$]$~$\lambda$3727 doublet will drop,
and the $[$\ion{O}{3}$]$~$\lambda$5007/$[$\ion{O}{2}$]$~$\lambda$3727
ratio will increase.
We find from the density and temperature sensitive
emission-line ratios described below, however, 
that the electron densities in the NLR are probably
lower than those conducive to collisional de-excitation.
In either case, with or without collisional de-excitation,
the drop in the ionization parameter sensitive diagnostics
with increasing distance from the nucleus indicates a radial
dependence on electron density.
Photoionization modeling will
be employed in Paper II to further investigate the physical
conditions that produce these observed gradients.

The line ratio $[$\ion{O}{3}$]~\lambda\lambda$(5007+4959)/$\lambda$4363
is a sensitive temperature diagnostic for a photoionized gas
in the density range 10~cm$^{-3}~<~n_{e}~<$~10$^{4}$~cm$^{-3}$.
The theoretical value of this
ratio decreases with increasing temperature. We detect no obvious
trend with spatial position for this diagnostic. The scatter in
the $[$\ion{O}{3}$]$ ratio corresponds to an uncertainty in the
temperature ranging from 12000~K to 17000~K.
A similar temperature range is found in the NLR of NGC~4151 \citep{nel00}.

\begin{figure}
\figurenum{14}
\begin{minipage}{6.25in}
\plotone{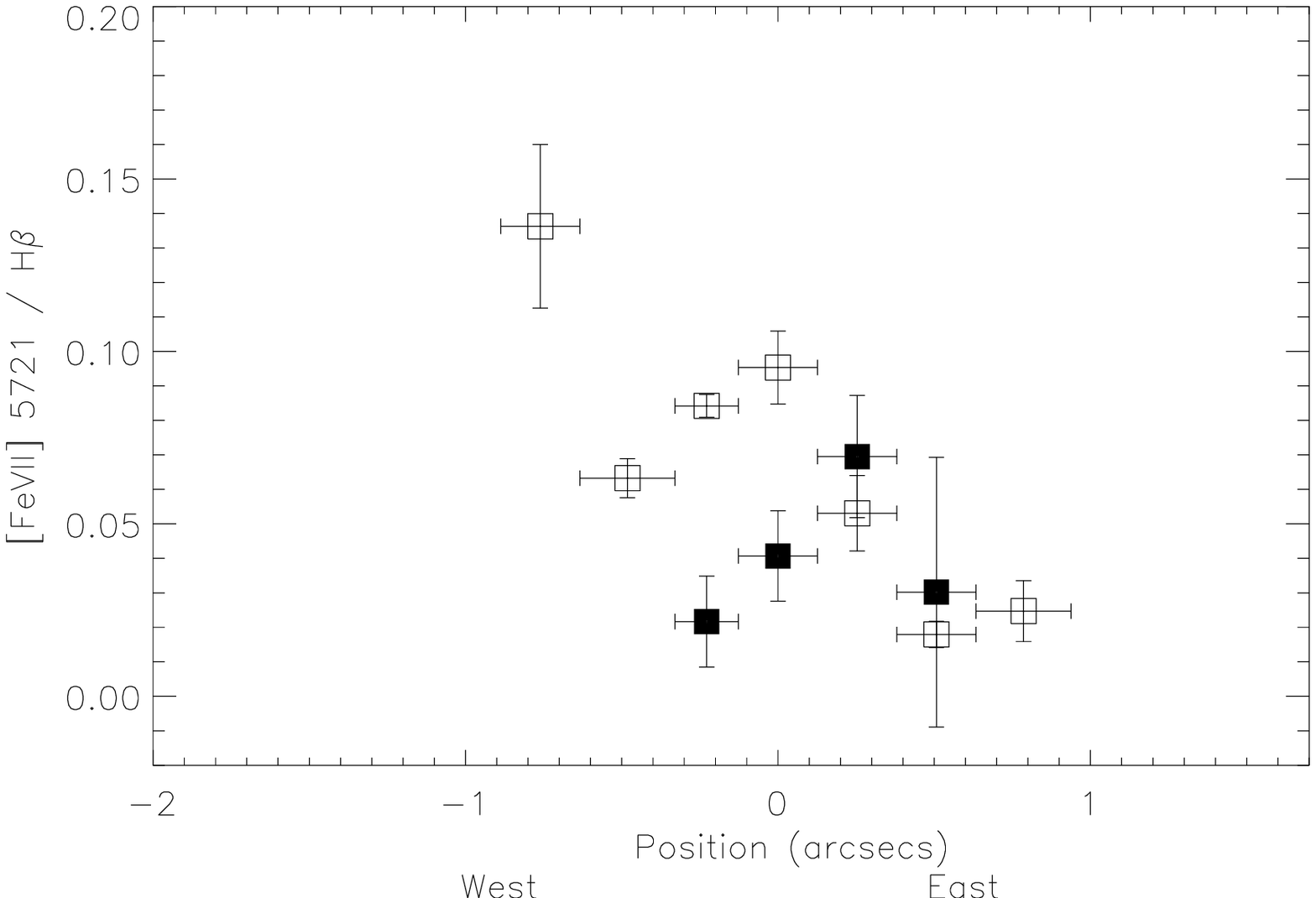}      
\caption{$[$\ion{Fe}{7}$]$~$\lambda$5721/H$\beta$~$\lambda$4861
plotted as a function of  angular separation from the nucleus along
the slit.  Note the steep drop in the ratio to the east.
} \label{fig13}
\end{minipage}
\end{figure}

\begin{figure}
\figurenum{15}
\begin{minipage}{6.25in}
\plotone{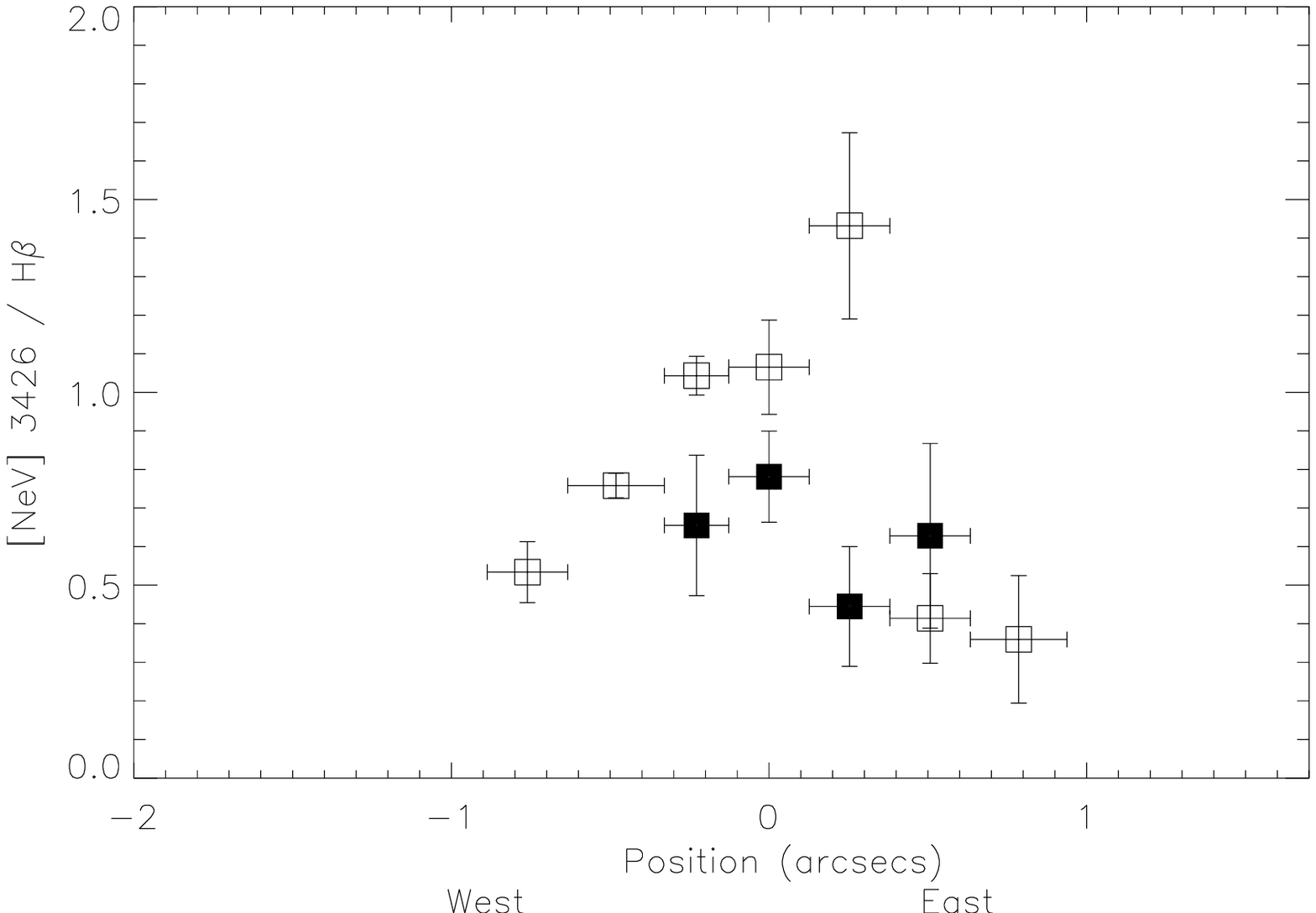}      
\caption{$[$\ion{Ne}{5}$]$~$\lambda$3426/H$\beta$~$\lambda$4861
plotted as a function of  angular separation from the nucleus along
the slit.  Note the steep drop in the ratio to the east.
} \label{fig12}
\end{minipage}
\end{figure}

Although the energy required to produce most of the ions observed in the
Markarian~3 NLR is less than 65~eV, two ions, Fe$^{+6}$ and
Ne$^{+4}$ require energies of 100~eV and 97~eV, respectively.
The spatial profiles of emission lines from these species exhibit
striking differences from the lower ionization lines at
neighboring wavelengths. The flux drops rapidly from the nucleus
to the eastern-most detectable component. This is most
dramatically illustrated by comparing the
$[$\ion{Fe}{7}$]$~$\lambda$5721 and $\lambda$6087 lines with their
neighboring lines,
$[$\ion{N}{2}$]$~$\lambda$5755 and \ion{He}{1}~$\lambda$5876, in
the mode G750L spectrum (Figures~\ref{slide9} and \ref{slide11}).
The fact that lines close in wavelength do not show the same
spatial morphology suggests that this is not a reddening effect.
The flux measurements for the $[$\ion{Fe}{7}$]$~$\lambda$5721 line
relative to H$\beta$~$\lambda$4861 are plotted in
Figure~\ref{fig13}.
Figure~\ref{fig12} shows the flux measurements of
$[$\ion{Ne}{5}$]$~$\lambda$3426 relative to H$\beta$~$\lambda$4861.
The asymmetric spatial profile of the emission-lines from
these high-ionization potential ionic species suggests that an
absorber located between the BLR and the NLR blocks the continuum
above $\sim$65~eV that would be required to ionize
Fe$^{+5}$ and Ne$^{+3}$ in the eastern side of the
NLR.  Similar effects have been suggested for NGC~4151 \citep{ale99}
and NGC~1068 \citep{ale00}.

\section{The Observed Continuum}

\subsection{Spectral Decomposition of the Continuum}

\begin{figure}
\figurenum{16}
\begin{minipage}{6.25in}
\plotone{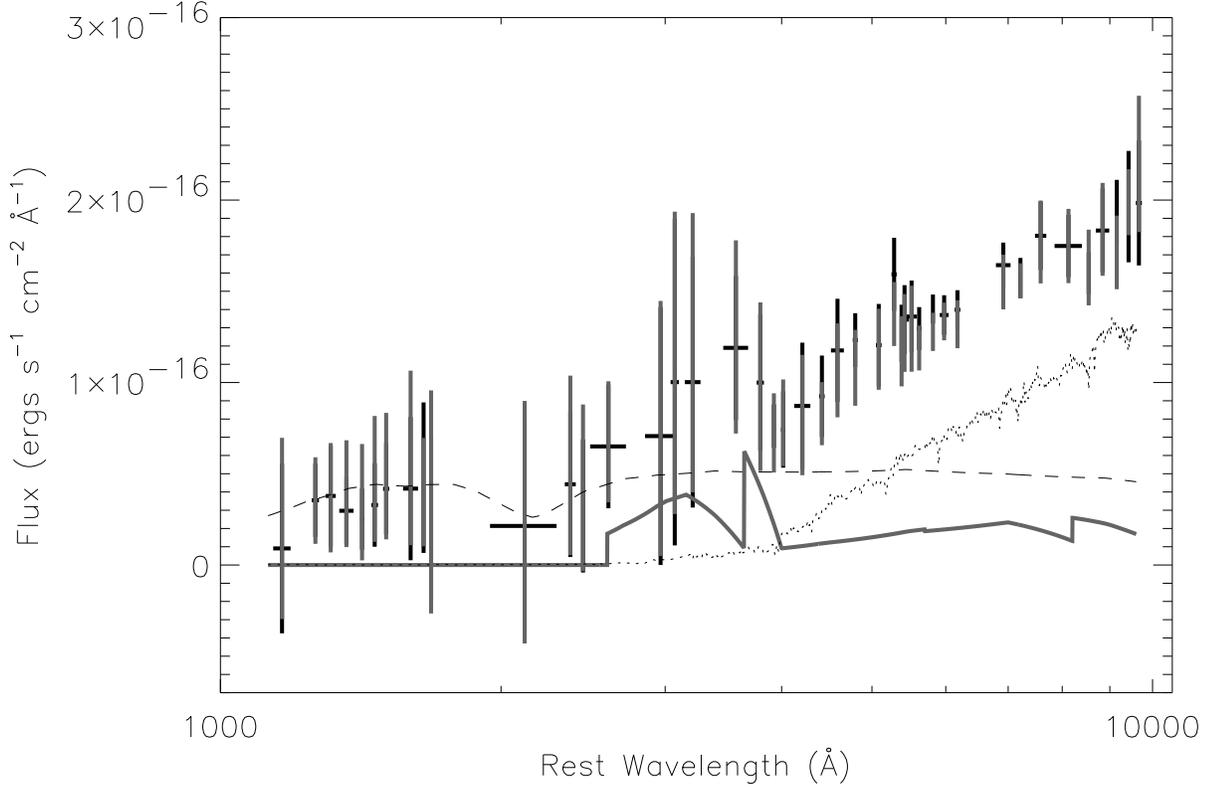}     
\caption{The  eastern and western extracted continuous spectra (crosses) 
and individual model components. 
Crosses denote average continuum 
values measured between emission lines: the horizontal bar is the width
of the measurement region in wavelength,
and the vertical bar is the $\pm$1$\sigma$ uncertainty
in the computed average.  
Eastern and western extractions are overplotted to 
emphasize their similarity after accounting for  
extinction and intrinsic luminosity. 
The gray crosses represent the eastern extraction
corresponding to the region encompassed by bins 
+0$\farcs$80 to +0$\farcs$30 in Figures~\ref{fig1}-\ref{fig12}.   
The black crosses represent
the western extraction (+0$\farcs$00 to -1$\farcs$00)
transformed to match the eastern extraction using (1) reddening
that corresponds to to E(B-V)=0.08 with the \citet{koo81} LMC
extinction curve and (2) a multiplicative scale factor of 1.4.
The dashed line is the power-law continuum ($\alpha$=1) 
reddened by an amount corresponding to E(B-V)=0.08 with 
the \citet{koo81} LMC extinction curve and E(B-V)=0.19 with 
the Galactic extinction curve of \citet{sav79}.  The dotted 
line represents the S0 galaxy template of \citet{kin96} 
reddened by E(B-V)=0.28 with the \citet{koo81} LMC extinction curve 
and by  E(B-V)=0.19 with the  Galactic curve \citep{sav79}. 
The solid gray line is the NLR gas recombination spectrum reddened 
in the same manner as the S0 template spectrum. 
The power-law and S0 galaxy template spectra were scaled 
so that f$_{galaxy}$=0.75 at 8125\AA~ and also so that 
the sum of these components with the recombination spectrum 
matches the data. 
A logarithmic wavelength scale,
adjusted to the rest-frame of Markarian~3
is used to separate the ultraviolet measurements.
} \label{continuumcomponentsfig}
\end{minipage}
\end{figure}

\begin{figure}
\figurenum{17}
\begin{minipage}{6.25in}
\plotone{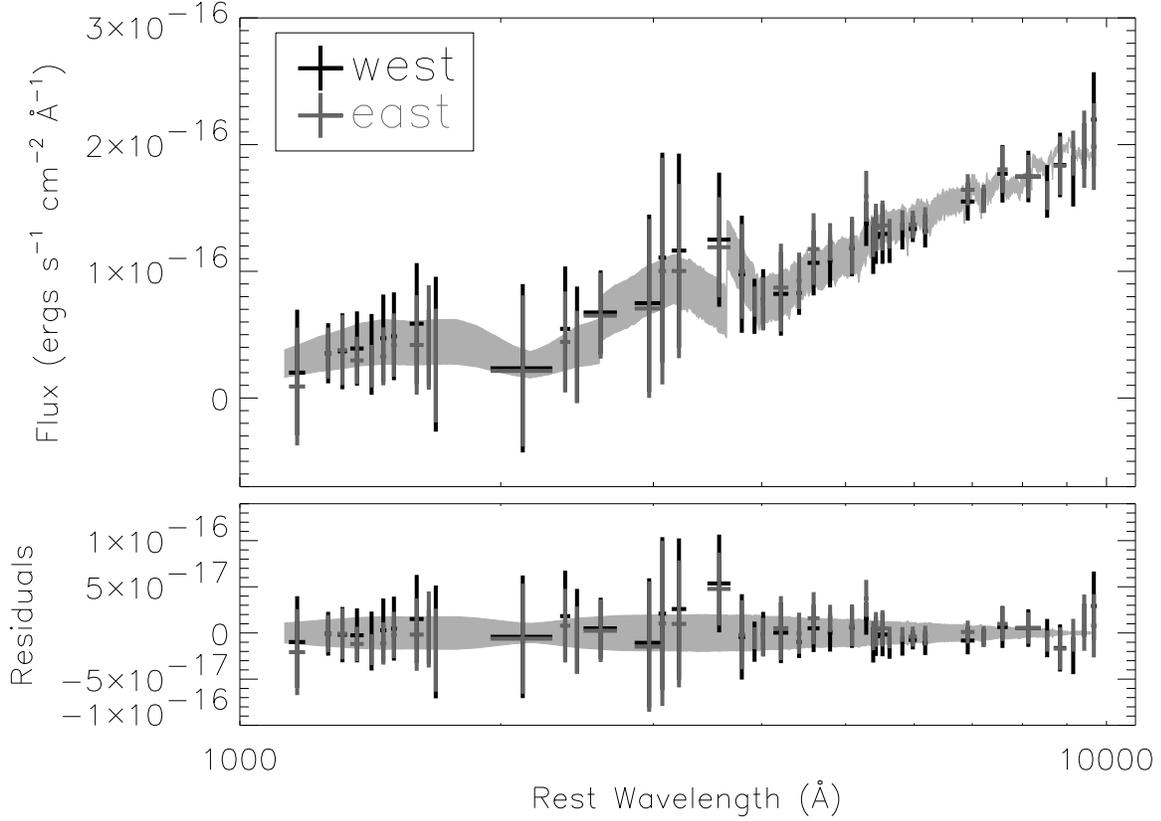}     
\caption{The  eastern and western extracted continuous spectra (crosses) 
and the best fit model (thick gray line).  
The extracted spectra are identical to those shown in 
Figure~\ref{continuumcomponentsfig}. 
The thick gray line shows the variation in the
reddened model spectrum (S0 galaxy + power-law + H$^{+}$ and He$^{+2}$
recombination continua, see text and Figure~\ref{continuumcomponentsfig} 
caption) when the galaxy
fraction (f$_{galaxy}$ = L$_{galaxy}$/(L$_{galaxy}$+L$_{scattered-AGN}$)  
is varied between 0.65 and 0.85 at 8125~\AA.
} \label{continuumfig}
\end{minipage}
\end{figure}

We examined the observed continuum in  the same
1$\farcs$8 region spanned by the bins defined
for the emission-line measurements.
Since we found a reddening gradient along the STIS slit
increasing from west to east from emission-line diagnostics,
we surmised that the intrinsic continuum may also be similarly
reddened.  We made separate continuum extractions for the
east and for the west.  The western extraction is the sum of 
the detector rows corresponding to the bins 
labeled -0$\farcs$8, -0$\farcs$5, -0$\farcs$2, and 
0$\farcs$0  in Tables 2 and 3.  The eastern extraction is the 
sum of the rows corresponding to bins 0$\farcs$3, 0$\farcs$5, 
and 0$\farcs$8.  For both extractions we exclude those regions 
containing emission lines in the row sum. 
After applying a foreground reddening corresponding to E(B-V)=0.08 
(LMC extinction curve [Koornneef \& Code 1981]) 
to the western extraction, we found that its shape matched that of 
the eastern extraction.  The flux level of the eastern 
extraction is 1.4 times greater than that of the extinction-transformed 
western extraction over the entire observed wavelength range. 
The eastern spectral extraction
and the extinction-transformed western extraction
are shown with the model components for the eastern extraction 
in Figures~\ref{continuumcomponentsfig}-\ref{continuumfig}.  
Note that the 
extinction-transformed western extraction is scaled by 
1.4 for comparison with the eastern extraction. 

We modeled the observed continuum as a combination of
host galaxy stellar light,  scattered light from the
active nucleus, and recombination emission from
H$^{+}$ and He$^{+2}$ in the NLR.
Stellar light solely from an old population has
insufficient far-UV (STIS mode G140L) flux
to match the observed spectrum.   We determined
the relative contribution of the stellar-light and scattered
AGN-light components to our observed
spectrum using the \citet{kin96} spectral templates for normal
galaxies and a simple power-law spectrum.
We used the \citet{kin96} S0 galaxy template for the normal galaxy
component.  We assumed that the  spectral index of the power-law component
in the observed spectral region is $\alpha$=1.
The recombination continuum was determined using
the recombination-line and continuous-emission calculations 
described in \citet{ost89}.  The input variables for these 
calculations were the sums of 
the extinction corrected H$\beta$~$\lambda$4861
and \ion{He}{2}~$\lambda$4686  measurements from Table~3 in those 
bins corresponding to eastern (+0$\farcs$3 to 0$\farcs$8) 
and western (-0$\farcs$8 to 0$\farcs$0) extractions.
We applied  the Galactic
extinction  (corresponding to E(B-V)=0.19 $[$Schlegel et al. 1998$]$)
in the direction of Markarian~3 to
all components of the model spectrum using the \citet{sav79}
extinction curve.  We also applied an extinction
corresponding to E(B-V)=0.08 using the LMC curve \citep{koo81}
to all components
of the model spectrum to match the eastern extraction.
We normalized the stellar and power-law components of the
model spectrum so that they
matched the data at the 8125~\AA~ measurement bin of the eastern extraction.
We scaled and added the two components together, varying the scale factor
to determine the ratio of host galaxy to scattered AGN light that best
matched the overall shape of our observed continuum.
The signal-to-noise is too low to permit matching absorption lines
between the data and the galaxy template.
The recombination continuum model components were added
to the power-law~+~host galaxy model.
A host galaxy to power-law light ratio of 3:1 at 8125~\AA~
yielded a good match to the data, but we found that additional extinction
intrinsic to the Markarian~3 host galaxy, was required to give the best
fit at all wavelengths.  Applying additional extinction corresponding to
E(B-V)=0.20 with the LMC curve \citep{koo81} to  the host
galaxy component and to the recombination continuum
components gave the best fit to the eastern extraction.
The reddened model components for the best fit to the eastern extraction 
are shown with the eastern and western extractions  
in Figure~\ref{continuumcomponentsfig}.   The sum of the three 
model components are shown with the data in Figure~\ref{continuumfig}.

The uncertainty in the host galaxy fraction
(f$_{galaxy}$ = L$_{galaxy}$/(L$_{galaxy}$+L$_{scattered-AGN}$, 
where L$_{galaxy}$ and L$_{scattered-AGN}$ are the host galaxy 
and power-law continua, respectively)
is shown by the gray shaded area in Figure~\ref{continuumfig}.
The galaxy fraction at 8125~\AA~ ranges from 0.65 to 0.85 across this region.
We estimated the uncertainty in the host galaxy extinction
for a given host galaxy template and host galaxy extinction
value for a fixed  galaxy fraction at 8125~\AA~ of  0.75.  Varying the
host galaxy color-excess from $\pm$0.05
matches the scatter in the smallest error bars at wavelengths
longer than 4000~\AA.   A variation of $\pm$0.10 matches the
scatter in the largest error bars.  Variations at these levels
have no effect on the model spectrum at wavelengths less than
4000~\AA, where the scattered continuum from the hidden AGN dominates.

\begin{figure}
\figurenum{18}
\begin{minipage}{6.25in}
\plotone{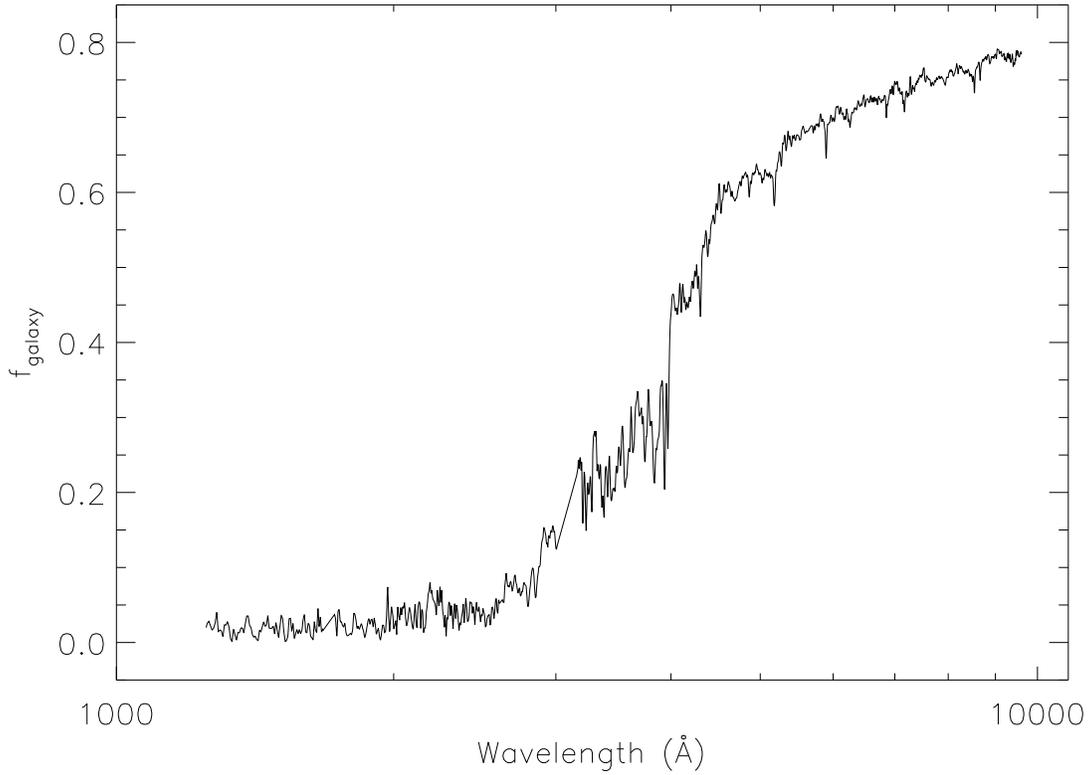}      
\caption{The galaxy fraction (f$_{galaxy}$), or the ratio of 
host galaxy continuum (\citet{kin96} S0 template) 
to the sum of the host galaxy continuum and  
power-law continuum ($\alpha$=1) as a function
of wavelength for the best fit to the observed continuum. 
f$_{galaxy}$ = L$_{galaxy}$ / (L$_{galaxy}$ + L$_{scattered-AGN}$) = 0.75 
at 8125~\AA. 
} \label{galfrac}
\end{minipage}
\end{figure}

Our results for continuum decomposition have a lower
galaxy fraction than the previous studies
of \citet{kos78}, \citet{mal83}, and \citet{tra95}.
The wavelength dependence of the galaxy fraction on
our continuum model normalized to 0.75 at 8125~\AA~ is
shown in Figure~\ref{galfrac}.
We estimated that the host galaxy fraction reaches 0.8 at the longest
wavelengths ($>$9500~\AA) in our model and drops to 0.65 near 4500~\AA.
The fraction  drops off sharply towards shorter wavelengths,
amounting to only 0.05 at 2600~\AA.  The fraction gently
decreases to 0.02 at far-UV wavelengths.
\citet{kos78} reported a galaxy fraction at 4861~\AA~ of 0.78.
\citet{mal83} found a galaxy fraction of 0.88 at 5175~\AA~
and \citet{tra95} computed a fraction of 0.88 at 5500~\AA.
We found galaxy fractions close to  0.60 at 4861~\AA~ and 5171~\AA,
and 0.70 at 5500~\AA.
These differences are  likely due to differences in aperture size.
The previous studies used apertures of width 2$\farcs$4, 1$\arcsec$, and
2$\farcs$4, respectively, which admit more host galaxy
light than the STIS 0$\farcs$1 wide slit.  If the same amount
of scattered nuclear continuum light is observed through all
apertures, the host galaxy fraction derived from a large aperture
observation will be higher than that derived from a small aperture
observation.

We found that the total flux of our observed continuum,
integrated in wavelength from 1164~\AA~ to 9670~\AA,
is 2$\times$10$^{-12}$~ergs~s$^{-1}$~cm$^{-2}$.
When corrected for Galactic extinction only, this value
is 3.5$\times$10$^{-12}$~ergs~s$^{-1}$~cm$^{-2}$.
At a distance of 53~Mpc, this  total luminosity is
1.1$\times$10$^{42}$~ergs~s$^{-1}$.

The scattered AGN
continuum luminosity can be derived from our power-law fit
by recalling that the western component has an unreddened power-law
shape with $\alpha$=1 and the eastern component is a reddened
(E(B-V)=0.08) version of the western component and is
1.4 times brighter.  When we add these two components together,
and integrate from 1164~\AA~ to 9670~\AA~ we find that the
luminosity is 5.8$\times$10$^{41}$~ergs~s$^{-1}$.
The total luminosity of the unreddened AGN continuum scattered light
integrated from  1164~\AA~ to 9670~\AA~ is
7.4$\times$10$^{41}$~ergs~s$^{-1}$.

\subsection{The Covering Factor}

We estimated the covering factor, $C$, of the NLR gas using the
relationship between H$\beta$ luminosity and ionizing continuum
given by \citet{net93}
\begin{equation}
L(H\beta) \simeq 7.4 \times 10^{13} \frac{L_{1}C}{\alpha} ergs~s^{-1}
\end{equation}
where $L_{1}$ is the monochromatic luminosity at 1~Ryd in
ergs~s$^{-1}$~Hz$^{-1}$ and $\alpha$ is the
power-law index above 1~Ryd.  This equation applies to an
optically thick gas under case B recombination conditions.
We summed our extinction-corrected
H$\beta$~$\lambda$4861 flux measurements listed in Table~\ref{tab:3}
then scaled by the distance factor $4 \pi D^{2}$ to find a total
H$\beta$~$\lambda$4861 luminosity of
7.2$\pm$0.4$\times$10$^{40}$~ergs~s$^{-1}$.
To find the luminosity at 1~Ryd, we assumed that the ionizing
continuum has the form of a two component power-law with a
break-point at $\nu$=(0.2~keV)/h, where h is Planck's constant.
We used the spectral
index of the ionizing continuum derived in \S4.3 of $\alpha$=1.36
for the range (13.6~eV/h)$<\nu<$(0.2~keV/h).
\citet{tur97} used the  \emph{ASCA} spectrum of Markarian~3
to estimate that the unabsorbed, intrinsic X-ray spectrum has a
spectral index $\alpha$=1 over the
range (0.2~keV/h)$<\nu<$(50~keV/h) and that its integrated flux from
2~keV to 10~keV is $\sim$10$^{44}$~ergs~s$^{-1}$.
Using this normalization, we found that
L$_{1}$=5.0$\times$10$^{28}$~ergs~s$^{-1}$~Hz$^{-1}$,
and the covering factor is 2.7$\pm$0.15\%.
\citet{net93} found typical NLR covering factors for samples of Seyfert 1
galaxies and radio-loud quasars to lie between 1\% and 4\%.
We note that the total ionizing continuum luminosity from 13.6~eV to
50~keV using this two-component model is 6.2$\times$10$^{44}$~ergs~s$^{-1}$.

The covering factor value derived from
the H$\beta$~$\lambda$4861  measurement should be a lower limit since
our equivalent aperture of size 0$\farcs$1 $\times$ 1$\farcs$8
does not sample the total emitted H$\beta$~$\lambda$4861 flux.
From purely geometrical considerations, the maximum possible covering factor
for our adopted hollow bi-cone geometry (with opening half angles of
15$\degr$ and 25$\degr$) is ($\Delta\Omega$/4$\pi$)$\times$100\% = 6\%.
At our adopted distance to Markarian~3 the slit width is 26~pc
and our H$\beta$ measurement region covers $\sim$30\% of the NLR bi-cone.
The upper limit to the
covering factor constrained by the superposition of the STIS slit
on our bi-cone model is 6\% $\times$ 30\% = 2\%.  This is on the order of
our photon-counting derivation lower-limit.  Based on this
analysis the covering factor within the slit is unity given
the bi-cone geometry.   However,
if we have underestimated the value of L$_{1}$, the covering
factor may actually be lower (see below).

We estimated the fraction of intrinsic
non-ionizing AGN continuum emission scattered into our line-of-sight.
We extrapolated our power-law fit of the observed continuum
($\alpha$=1) to $\nu$=(13.6~eV)/h and scaled it to the value
of $L_{1}$ estimated for the ionizing continuum.
The inverse of this scale factor, 0.2\%, is the scattering fraction.

We compared the scattering fraction of the continuum with
that for BLR emission.
We detected scattered \ion{C}{4}~$\lambda\lambda$1548,1551 broad-line emission
in a  large bin (1$\farcs$8) extraction equivalent to the full spatial
range of the seven emission-line measurement bins.
These broad-line wings are indicated by the heavy gray line
in Figure~\ref{slide6}.
Other permitted lines in the STIS spectrum
may have scattered broad components, but they also have neighboring
lines which make such identification difficult.  Of the permitted lines
observed in the STIS spectrum, \ion{C}{4}~$\lambda\lambda$1548,1551
and \ion{Mg}{2}~$\lambda\lambda$2796,2803
are the most isolated.  We do not observe scattered broad-line
flux from \ion{Mg}{2}.

We followed the procedure of \S3.1 to measure the fluxes of the various
components of the blended \ion{C}{4}~$\lambda\lambda$1548,1551 lines.
We assumed that each narrow line consists of two kinematic components:
one redshifted and the other blueshifted.  We identified the
individual narrow line widths and velocities, relative to the systemic
velocity, by using the $[$\ion{O}{3}$]~\lambda5007$ line as a
template.  We fitted the remaining flux above the continuum with a
Gaussian function.  The scattered broad-line component flux is
1.25$\times$10$^{-14}~$ergs~s$^{-1}$~cm$^{-2}$.  The broad-line
FWHM of 24~\AA~ corresponds to a velocity dispersion of
$\pm$2200~km/s.  After correcting for Galactic extinction \citep{sav79}
(corresponding to E(B-V)=0.19, $[$\citet{sch98}$]$) we found a flux of
5.2$\times$10$^{-14}~$ergs~s$^{-1}$~cm$^{-2}$.  Accounting for
the 53~Mpc distance to Markarian~3 the luminosity of the scattered
line emission is 1.6$\times$10$^{40}~$ergs~s$^{-1}$.
We note that the equivalent width of the broad-line component
is $\sim$125\AA.  This is consistent with the measurements
of \citet{wu83} for Sy~1 galaxies.

We estimated the intrinsic \ion{C}{4}~$\lambda\lambda$1548,1551
luminosity by assuming a fixed ratio of hard X-ray flux to
broad-line flux.  We scaled the Markarian~3 hard X-ray flux by that ratio
determined for Fairall~9, a Sy~1 galaxy with relatively unabsorbed AGN
X-ray continuum emission.
We obtained the hard X-ray flux, f$_{2-10~keV}$=2$\times$10$^{-11}$
ergs~s$^{-1}$~cm$^{-2}$, from the TARTARUS database.  \citet{wu83}
listed a \ion{C}{4}~$\lambda\lambda$1548,1551 flux of 7.31$\times$10$^{-12}$
ergs~s$^{-1}$~cm$^{-2}$ for Markarian~3.  
Scaling our adopted hard X-ray luminosity
of 10$^{44}$~ergs~s$^{-1}$ for Markarian~3 by the ratio of
\ion{C}{4}~$\lambda\lambda$1548,1551 to hard X-ray flux for
Fairall~9 yielded an estimated \ion{C}{4}~$\lambda\lambda$1548,1551
luminosity of 3.7$\times$10$^{43}$ ergs~s$^{-1}$ for Markarian~3.
Dividing our measurement of the scattered line luminosity by
this estimated intrinsic luminosity indicates that the
line emission scattering fraction is 0.04\%.  This is a
factor of 5 less than our estimate for the continuum scattering fraction.

We can obtain a better agreement between the continuum and line emission
scattering fractions if the ionizing continuum has a steeper
spectral index for the range (13.6~eV/h)$<\nu<$(0.2~keV/h).
A larger value for L$_{1}$ will lower
the scattering fraction for the non-ionizing continuum emission.
We hold the hard X-ray normalization fixed at 10$^{44}$ ergs~s$^{-1}$
with $\alpha$=1.    We used an X-ray to optical energy index
of $\alpha_{OX}$=1.5 \citep{tur97} to predict the intrinsic luminosity
at 2500~\AA.  This energy index is the modal value of the
$\alpha_{OX}$ distribution for the
Braccesi BF quasar subsample studied by \citet{wil94}.
We used the
non-ionizing continuum index of $\alpha$=1 to extrapolate
from 2500~\AA~ to 912~\AA (13.6~eV).
We then found that the spectral index for the range
(13.6~eV/h)$<\nu<$(0.2~keV/h) is $\alpha\sim$2.
This is the same index that \citet{kra86} used
to model the Markarian~3 NLR emission lines.
The new value of L$_{1}$ is 2.78$\times$10$^{29}$~ergs~s$^{-1}$.
Now the scattered fraction of continuum emission is
0.04\%, which is consistent with the scattered fraction of intrinsic
line emission.  We note that the new values of spectral index
and ionizing luminosity at 1~Ryd yield a covering factor
for the NLR gas of $>$0.7$\pm$0.04\%.  This value is well
below the geometrical upper limit of 2\%. The integrated
ionizing luminosity from 13.6~eV to 50~keV is now
1.2$\times$10$^{45}$ ergs~s$^{-1}$.

The discrepancy between the spectral index derived from the
\ion{He}{2}~$\lambda4686$/H$\beta$ ratio and that
derived from extrapolation using the X-ray to optical
energy  index may indicate that
the Markarian~3 NLR contains some combination of matter-bounded
and ionization-bounded clouds.
The former derivation applies only to ionization-bounded clouds.
Additionally, the ionizing continuum
might not be well characterized by a simple power law.  It may
have a ``big blue bump'' in the EUV (\citet{sun89}; \citet{kro91}).
We will investigate these possibilities using detailed photoionization
models in Paper II.

\section{Orientation of the NLR Bi-Cone With Respect to the Host Galaxy}

\citet{sch00} observed an ellipticity (e=1-b/a) of 0.159 in a
60~$\mu$m image of Markarian~3.  This corresponds to an inclination
of 33$\degr$ ($i = arccos(b/a)$).  However, they were unable to determine the
orientation of the Markarian~3 galactic disk.
Galactic disk orientation may be determined using 
rotation curves derived from neutral hydrogen (21~cm) 
observations, but no such information for Markarian~3 is currently 
available in the literature.

\begin{figure}
\figurenum{19}
\begin{minipage}{6.25in}
\plotone{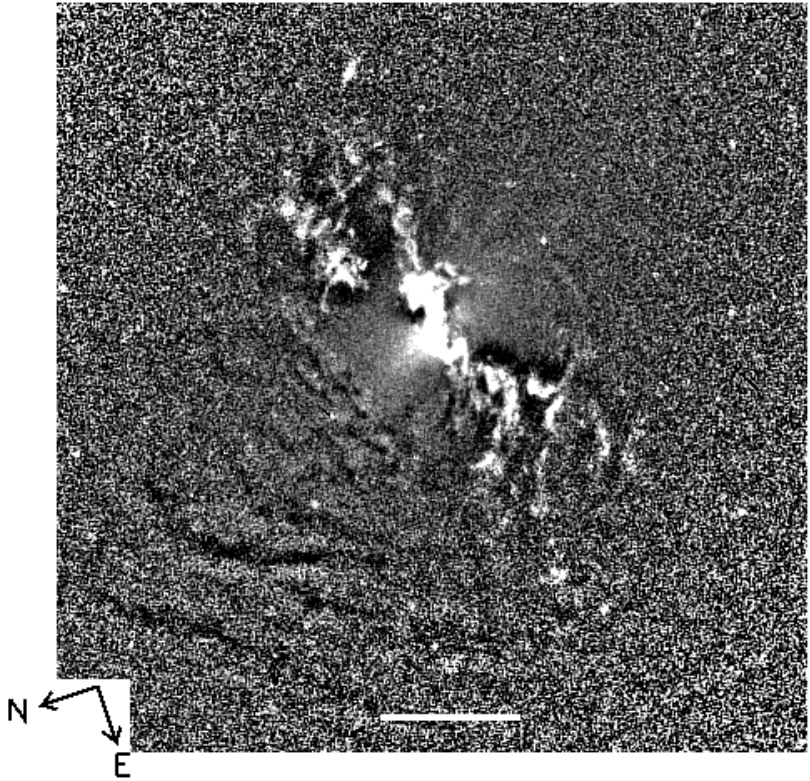}      
\caption{Unsharp-masked version of archival HST/WFPC2/F606W image.  The
unsharp-mask filter size is 1$\farcs$43.  The horizontal bar corresponds
to 3$\farcs$8 or 1~kpc.
The display scaling is such that emitting-material appears white and
obscuring materials (i.e., dust lanes) appear black.
Note the presence of dust lanes towards the north-east of the NLR and
lack of dust lanes to the south-west.  The emission in the central region
is primarily due to emission-lines from the  NLR
within the F606W bandpass.
} \label{slide2d}
\end{minipage}
\end{figure}

\citet{sch00} suggest that dust lanes should lie only on the
closer side of a galactic disk, as they would be delineated by
background bulge light.  We searched for dust lanes in an archival
HST/WFCPC2/F606W image obtained by \citet{mal98} (exposure
time is 500~seconds). After
cleaning the cosmic-rays on the single archival image, we created an
unsharp-masked image of Markarian~3 to enhance high contrast
features that are overwhelmed by the bulge light of Markarian~3.
The unsharp-masking process requires dividing an image by a median
filtered copy of the same image.  We used a 31$\times$31~pixels squared  
(1$\farcs$43$\times$1$\farcs$43) two-dimensional  median filter.  
The unsharp-masked F606W image is shown in
Figure~\ref{slide2d}.  The display in Figure~\ref{slide2d} is such
that emitting-material appears white and
obscuring material (i.e., dust lanes) appear black.
The horizontal bar corresponds to 1~kpc or 3$\farcs$8.
Note the preponderance
of concentric dust lanes found between 4$\arcsec$ and 12$\arcsec$ north-east
of the nucleus, and the lack of such lanes in the south-west.
This is consistent with the reddening gradient observed 
along the STIS slit that increases from west to east as 
shown in Figure~\ref{fig1}.
Based on the orientation of the galaxy disk major axis
(P.A.=28$\degr$ east of north [Schmitt and Kinney 2000]) 
and using the bulge back-lighting
assumption, we would expect such arcs to be oriented perpendicular
to the minor axis on the near side of the galaxy, not  
perpendicular to the major axis as they appear in 
Figure~\ref{slide2d}. 
However, the presence of dust lanes in the east and not in the west
and the bulge backlighting hypothesis suggest that the 
host galaxy disk is tilted towards the observer in the east along 
the STIS slit (P.A.=70$\degr$ [Schmitt and Kinney 1996]). 

We used these clues to the host galaxy inclination and the
results of our line diagnostic and continuum
studies to develop a physical
picture of the Markarian~3 NLR.  The reddening gradient derived from the
\ion{He}{2}~$\lambda\lambda$1640,4686 line emission
shown in Figure~\ref{fig1}
suggests the presence of a foreground screen between the observer
and the NLR, with increasing extinction from west to east.
An extinction gradient is supported by the continuum observations
as well:  the continuum shape of the western extraction
matches the eastern extraction only after applying an amount
of extinction corresponding to E(B-V)=0.08 using
the \citet{koo81} LMC curve. We infer
that the foreground screen is  dust within the inclined disk of the host
galaxy and  argue that the plane of the host galaxy is tilted towards
us in the east.  We adopt the disk inclination of 33$\degr$ from
\citet{sch00}.  \citet{rui01} showed that the NLR bi-cone is
tilted towards us in the east as well, although at a
smaller angle (5$\degr$).  The eastern bi-cone is more heavily reddened
because we view it through the dust within the galactic plane. The
western bi-cone is less reddened since it lies above the plane of
the galaxy, or at least the line-of-sight intersects less planar
dust than that towards the eastern bi-cone.  This configuration
is consistent with the findings of \citet{elm99}. They suggest
that the near side of an inclined disk galaxy will be redder than the
far side because the light source is behind most of the disk dust.

In addition to
reddening in the east, the contrast in the extinction
gradient might be enhanced by
preferentially scattered blue continuum emission in the west.
If the host galaxy disk is tilted away from
us in the west, dust within the disk and throughout or behind the
NLR might scatter blue continuum emission into
our line-of-sight.  \citet{kis02} found evidence of more
scattering (possibly due to dust) in the west than in
in the east from a polarized-flux color-map (HST/WFPC2 F275W-F342W)
of the Markarian~3 NLR.

\begin{figure}
\figurenum{20}
\begin{minipage}{6.25in}
\plotone{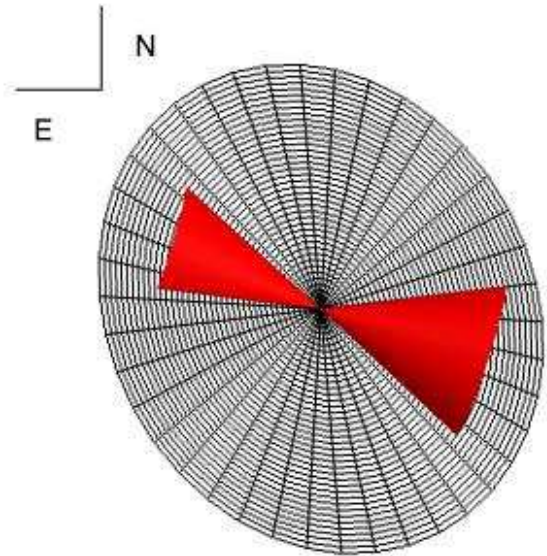}      
\caption{Schematic of the Markarian~3 NLR and its relative
orientation to the host galaxy disk
from the observer's point-of-view. North is up and East is to the left.}
\label{obs_cartoon}
\end{minipage}
\end{figure}

\begin{figure}
\figurenum{21}
\begin{minipage}{6.25in}
\plotone{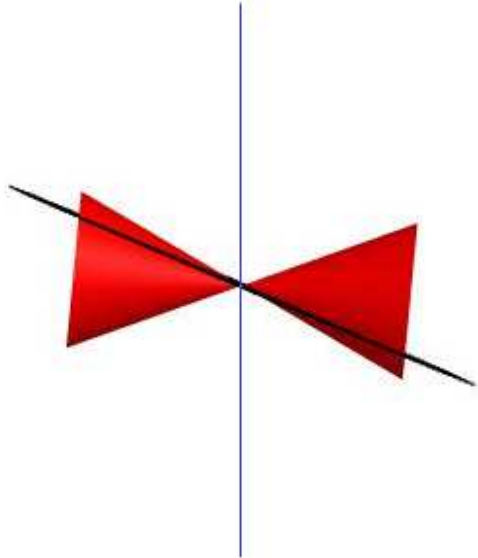}      
\caption{Schematic of the Markarian~3 NLR and its relative
orientation to the host galaxy disk, which is edge-on in this figure.
The projected line-of-sight is from the top of the figure and
parallel to the vertical bar.
East is to the left of the figure.}
\label{perpendicular_cartoon}
\end{minipage}
\end{figure}

Figures~\ref{obs_cartoon} and \ref{perpendicular_cartoon} are
schematic diagrams  of our physical picture of the Markarian~3 NLR
shown from two different
viewing angles.  We show the relative orientation
of the components, not their relative spatial scales.
The host galaxy disk position-angle measured east
from north is 28$\degr$ and its inclination is 33$\degr$ \citep{sch00}.
The host galaxy disk is tilted towards the observer in the east.
The bi-cone orientation parameters from \citet{rui01} are:
position-angle (measured east from north) = 72$\degr$,
outer-edge opening half-angle = 25$\degr$, and
inclination = 5$\degr$ out of the plane of the sky towards
the observer in the east. These parameters are derived from the
$[$\ion{O}{3}$]$~$\lambda$5007 morphology and kinematics  of the
inner 2-3~$\arcsec$ of the NLR.  In Figure~\ref{obs_cartoon}
the point-of-view is that of the observer
such  that the plane of the page is parallel to the plane of the sky.
North is at the top of the figure and
east is to the left.  In Figure~\ref{perpendicular_cartoon}.  The vertical
line is parallel to the observer's  projected line-of-sight.
The observer views the system from the top of the figure.
We assume
from Figure~8 of \citet{rui01} that most of the NLR gas is confined
within the inner (15$\degr$ $[$Ruiz et al. 2001$]$ not shown in the figures)
and outer half-opening angles of the bi-cones.

\citet{rui01} found that there were more redshifted points in the
East further out in the ENLR than in the West at comparable
distances from the nucleus (see their Figure~4).  If the motions
of the ENLR are governed by the galactic disk rotation, these
observations and the inference that the disk is inclined towards
the observer in the East would be consistent with a clockwise
(north-to-west) disk rotation as viewed by the observer.
\citet{kot97} presented  color maps of Markarian~3.  Their V-R
color map shows a red spiral feature extending approximately
$\pm$4$\farcs$5 from the nucleus. If this feature has a kinematic
origin and if the spiral arms are trailing the rotation of the
galaxy, it would support the hypothesis that the galaxy rotation
runs clockwise from north to west.

\section{Summary}

We used HST/STIS low-resolution longslit spectra to study the
physical conditions of the Markarian~3 NLR.  We found an
extinction gradient in the emission-line and continuum flux increasing
from west to east along the STIS slit.  The extinction is best
characterized with an LMC-type \citet{koo81} curve.  We developed a
physical picture of the NLR and host galaxy geometry using
the extinction gradient, the NLR-gas kinematic model of \citet{rui01},
and the host galaxy inclination  of \citet{sch00}.
We infer that the host galaxy disk is tilted towards us in the
east, providing a dustier line-of-sight to the eastern NLR
bi-cone than that to the western bi-cone.

Using emission-line diagnostics for the dimensionless
ionization parameter (U) we found evidence for a large scale gradient
in electron density in the NLR. The density appears to decrease with radius,
although slower than (1/r)$^{2}$.   Small scale deviations from the 
trend indicate that the gas distribution within the NLR may 
be patchy.  Diagnostics for temperature show no
trend with position along the STIS slit.  The scatter in the diagnostic
suggests an uncertainty in temperature ranging from 12,000~K to
17,000~K.
Photoionization diagnostics are consistent with
AGN power-law ionizing radiation and show no evidence
for a starburst in Markarian~3.  We found evidence
that the NLR nitrogen to oxygen abundance ratio is a factor of 1.6 above the
solar value.

We decomposed the observed continuum as S0 host galaxy stellar light,
scattered AGN power-law continuum ($\alpha$=1), and recombination continua
from H$^{+}$ and He$^{+2}$.  The ratio of
host galaxy continuum to scattered power-law continuum is 3:1
at 8125~\AA~ in our aperture.
We derived the ionizing continuum (E$\geq$13.6~eV)
as a two-step power law with a breakpoint at 0.2~keV.
We used  the \citet{tur97} estimate of the unabsorbed  X-ray spectrum
above the breakpoint,
where $\alpha$=1 and the 2-10~keV integrated luminosity is
10$^{44}$~ergs~s$^{-1}$.
We found $\alpha$=2 for the spectral range 13.6~eV~$<$~E~$<$~0.2~keV. 
When we extrapolated from the ionizing continuum to lower energies 
(E~$<$~13.6~eV) and compared the result with the scattered-AGN 
component of our observed continuum fit we found that 
the observed scattered-AGN continuum in the UV is 0.04\% of the intrinsic 
AGN continuum.  This is consistent with our estimate for the 
percentage of intrinsic \ion{C}{4}$~\lambda\lambda$1548,1551 
line emission scattered into our line-of-sight. 
Given our adopted bi-cone
geometry, the STIS aperture through which we observed the
NLR (52$\arcsec \times$0$\farcs$1), and the ionizing continuum
slope $\alpha$=2 at 13.6~eV, the covering factor of the NLR
gas is 0.7\%$<$C$<$2\%.

In Paper II we will use the physical parameters derived from the
emission line diagnostics and the inferred ionizing continuum as
input to the photoionization model code CLOUDY \citep{fer88} to
confirm our physical picture of the NLR.  These models will put
tighter constraints on sizes, densities, and spatial distribution
of gas clouds in the NLR.

\acknowledgements

We thank Don Lindler for advice and assistance on using the
CALSTIS data calibration pipeline.  We are grateful to Charles Bowers and
Kazunori Ishibashi for offering valuable insight to co-aligning
the STIS spectral images.
We thank Eliot Malumuth and Philip Plait for help
investigating the fringing in STIS mode G750L.
Thanks also to Gary Bower for helpful discussions
regarding the host galaxy orientation.

This research has made use of the
NASA/IPAC Extra-galactic Database operated by the Jet Propulsion
Laboratory, and NASA's Astrophysics Data System Bibliographic Services.
We also made use of the TARTARUS database,
which is supported by Jane Turner and Kirpal Nandra under
NASA grants NAG5-7385 and NAG5-7067.

Some of the data presented in this paper were obtained from the
Multimission Archive at the Space Telescope Science Institute
(MAST). STScI is operated by the Association of Universities for
Research in Astronomy, Inc., under NASA contract NAS5-26555. Support
for MAST for non-HST data is provided by the NASA Office of Space
Science via grant NAG5-7584 and by other grants and contracts.

We acknowledge the financial support of NAG5-4103 and 
NAG5-13109.

\clearpage


\clearpage


\clearpage


\clearpage

\clearpage


\clearpage


\clearpage


\clearpage


\clearpage


\clearpage


\clearpage


\clearpage


\clearpage


\clearpage


\clearpage


\begin{deluxetable}{lllrrrrlr}
\rotate
\tablecolumns{9}

\tabletypesize{\scriptsize}
\tablewidth{0pc}
\tablenum{1}
\tablecaption{STIS Long Slit Observations of
Markarian 3\tablenotemark{a}\label{tab:1}}
\tablehead{
\colhead{Data Set} &
\colhead{Detector} &
\colhead{Optical} &
\colhead{Central} &
\colhead{Dispersion} &
\colhead{Spectral} &
\colhead{Plate} &
\colhead{Aperture} &
\colhead{Exposure}\\

\colhead{Set} &
\colhead{} &
\colhead{Element} &
\colhead{Wavelength} &
\colhead{Scale} &
\colhead{Resolution\tablenotemark{b}} &
\colhead{Scale} &
\colhead{} &
\colhead{Time}\\

\colhead{} &
\colhead{} &
\colhead{} &
\colhead{(\AA)} &
\colhead{(\AA/pix)} &
\colhead{(\AA)} &
\colhead{($\arcsec$/pix.)} &
\colhead{($\arcsec$)} &
\colhead{(seconds)}\\

\colhead{(1)} &
\colhead{(2)} &
\colhead{(3)} &
\colhead{(4)} &
\colhead{(5)} &
\colhead{(6)} &
\colhead{(7)} &
\colhead{(8)} &
\colhead{(9)}
}
\startdata
O5KS01J6Q & CCD      & MIRVIS & 7229 & N/A  & N/A  & 0.050 & F28X50LP  &  3 \tablenotemark{c}\\
O5KS01010 & CCD      & G430L  & 4300 & 2.73 & 4.00 & 0.050 & 52X0.1    &  1080\\
O5KS01020 & CCD      & G750L  & 7751 & 4.92 & 9.89 & 0.050 & 52X0.1    &  1080\\
O5KS01040 & FUV-MAMA & G140L  & 1425 & 0.60 & 1.13 & 0.025 & 52X0.1    &  3238\\
O5KS01050 & FUV-MAMA & G140L  & 1425 & 0.60 & 1.13 & 0.025 & 52X0.1    &  3238\\
O5KS01060 & NUV-MAMA & G230L  & 2376 & 1.58 & 2.35 & 0.025 & 52X0.1    &  3238\\
\enddata
\tablenotetext{a}{Observations were obtained under HST proposal
identification number 8480 by Kraemer and Crenshaw.}
\tablenotetext{b}{Spectral resolution is full-width at half-maximum (FWHM) in \AA~ and  is
related to $\sigma$ (for a Gaussian distribution) by
$\sigma$ = FWHM / 2.354.}
\tablenotetext{c}{Exposure time listed for O5K5S01J6Q is that for the third
exposure in the data set.  The same exposure is shown in
Figure~1.}
\end{deluxetable}


\newpage
%
%
\begin{deluxetable}{lrrrrrrrrrrrr}
\tabletypesize{\footnotesize}
\rotate
\tablewidth{530pt}
\tablenum{2}
\tablecolumns{13}
\tablecaption{Observed Emission-line Fluxes Relative to H$\beta$\label{tab:2}}
\tablehead{
\colhead{} &
\colhead{} &
\multicolumn{6}{c}{West} &
\multicolumn{5}{c}{East} \\
\colhead{} &
\colhead{} &
\colhead{   -0.8\tablenotemark{a}} &
\colhead{   -0.5} &
\colhead{   -0.2} &
\colhead{   -0.2} &
\colhead{    0.0} &
\colhead{    0.0} &
\colhead{    0.3} &
\colhead{    0.3} &
\colhead{    0.5} &
\colhead{    0.5} &
\colhead{    0.8} \\
\colhead{} &
\colhead{} &
\colhead{r\tablenotemark{b}} &
\colhead{r} &
\colhead{r} &
\colhead{b} &
\colhead{r} &
\colhead{b} &
\colhead{r} &
\colhead{b} &
\colhead{r} &
\colhead{b} &
\colhead{r} \\
\colhead{} &
\colhead{} &
\colhead{    0.3\tablenotemark{c}} &
\colhead{    0.3} &
\colhead{    0.2} &
\colhead{    0.2} &
\colhead{    0.3} &
\colhead{    0.3} &
\colhead{    0.3} &
\colhead{    0.3} &
\colhead{    0.3} &
\colhead{    0.3} &
\colhead{    0.3} \\
\colhead{(1)} &
\colhead{(2)} &
\colhead{(3)} &
\colhead{(4)} &
\colhead{(5)} &
\colhead{(6)} &
\colhead{(7)} &
\colhead{(8)} &
\colhead{(9)} &
\colhead{(10)} &
\colhead{(11)} &
\colhead{(12)} &
\colhead{(13)}
}

\startdata

Ly$\alpha$ & 1216 &      4.89 &      2.01 &      1.73 &      1.22 &      1.52 &      0.85 &      0.50 &      0.30 &      0.78 &      0.32 &      0.78 \\
 &  & (  0.16)\tablenotemark{d} & (  0.04) & (  0.02) & (  0.05) & (  0.05) & (  0.03) & (  0.07) & (  0.05) & (  0.05) & (  0.02) & (  0.02) \\
NV       & 1240 &      0.11 &      0.10 &      0.09 &      0.05 &      0.07 &      0.03 &      0.03 &      0.02 &      0.01 & $<$  0.02\tablenotemark{e} &      0.01 \\
 &  & (  0.01) & (  0.04) & (  0.00) & (  0.01) & (  0.00) & (  0.00) & (  0.00) & (  0.00) & (  0.00) & (  0.00) & (  0.01) \\
CII      & 1335 &      0.07 &      0.02 &      0.01 &      0.02 & $<$  0.02 & $<$  0.02 & $<$  0.06 &      0.00 &      0.00 &      0.00 &      0.02 \\
 &  & (  0.01) & (  0.01) & (  0.00) & (  0.00) & (  0.00) & (  0.00) & (  0.00) & (  0.00) & (  0.00) & (  0.00) & (  0.03) \\
NIV$]$     & 1486 &      0.06 &      0.07 &      0.04 &      0.10 &      0.02 &      0.04 & $<$  0.02 &      0.02 & $<$  0.01 &      0.01 &     ..... \\
 &  & (  0.07) & (  0.00) & (  0.00) & (  0.00) & (  0.00) & (  0.00) & (  0.00) & (  0.00) & (  0.00) & (  0.00) & (.....) \\
CIV      & 1548 &      0.75 &      0.62 &      0.81 &      0.31 &      0.49 &      0.22 &      0.15 &      0.03 &      0.14 &      0.03 &      0.07 \\
 &  & (  0.02) & (  0.04) & (  0.04) & (  0.01) & (  0.01) & (  0.02) & (  0.02) & (  0.01) & (  0.01) & (  0.02) & (  0.00) \\
CIV      & 1551 &      0.37 &      0.31 &      0.41 &      0.15 &      0.25 &      0.11 &      0.07 &      0.01 &      0.07 &      0.02 &      0.04 \\
 &  & (  0.01) & (  0.02) & (  0.02) & (  0.00) & (  0.01) & (  0.01) & (  0.01) & (  0.01) & (  0.00) & (  0.01) & (  0.00) \\
HeII     & 1640 &      0.41 &      0.35 &      0.38 &      0.20 &      0.22 &      0.19 &      0.10 &      0.12 &      0.13 &      0.12 &      0.14 \\
 &  & (  0.02) & (  0.01) & (  0.00) & (  0.01) & (  0.01) & (  0.00) & (  0.01) & (  0.02) & (  0.01) & (  0.01) & (  0.00) \\
OIII$]$    & 1665 &      0.23 &      0.04 &      0.01 &      0.02 &      0.01 &      0.10 &    ..... &      0.01 & $<$  0.01 &      0.07 &      0.03 \\
 &  & (  0.02) & (  0.01) & (  0.01) & (  0.02) & (  0.01) & (  0.01) & (.....) & (  0.00) & ( 0.00) & (  0.00) & (  0.00) \\
CIII$]$    & 1909 &      1.16 &      0.46 &      0.37 &      0.51 &      0.09 &      0.04 &      0.29 &      0.10 &      0.10 &      0.20 &      0.31 \\
 &  & (  1.62) & (  0.02) & (  0.00) & (  0.01) & (  0.01) & (  0.00) & (  0.04) & (  0.03) & (  0.01) & (  0.03) & (  0.07) \\
CII$]$     & 2324 &      0.44 &      0.15 &      0.13 &      0.16 &      0.11 &      0.11 &      0.09 &      0.10 &      0.13 &      0.10 &      0.21 \\
 &  & (  0.55) & (  0.00) & (  0.00) & (  0.00) & (  0.00) & (  0.00) & (  0.02) & (  0.01) & (  0.01) & (  0.01) & (  0.01) \\
$[$NeIV$]$   & 2424 &      0.25 &      0.24 &      0.26 &      0.15 &      0.18 &      0.10 &      0.14 &      0.05 &      0.08 &      0.06 &      0.03 \\
 &  & (  0.04) & (  0.02) & (  0.00) & (  0.01) & (  0.01) & (  0.01) & (  0.02) & (  0.01) & (  0.01) & (  0.00) & (  0.02) \\
$[$OII$]$    & 2470 & ..... &      0.09 &      0.05 &      0.03 &      0.02 &      0.05 &      0.07 & $<$  0.03 &      0.05 &      0.04 &      0.08 \\
 &  & (.....) & ( 0.00) & (  0.00) & (  0.01) & (  0.00) & (  0.01) & (  0.01) & (  0.00) & (  0.00) & (  0.00) & (  0.01) \\
MgII     & 2796 &      0.62 &      0.28 &      0.14 &      0.32 &      0.27 &      0.19 &      0.11 &      0.10 &      0.23 &      0.17 &      0.21 \\
 &  & (  0.02) & (  0.01) & (  0.01) & (  0.04) & (  0.01) & (  0.01) & (  0.02) & (  0.01) & (  0.02) & (  0.02) & (  0.04) \\
MgII     & 2803 &      0.31 &      0.14 &      0.07 &      0.16 &      0.14 &      0.09 &      0.06 &      0.05 &      0.12 &      0.08 &      0.10 \\
 &  & (  0.01) & (  0.01) & (  0.01) & (  0.02) & (  0.01) & (  0.01) & (  0.01) & (  0.01) & (  0.01) & (  0.01) & (  0.02) \\
$[$OIII$]$   & 3133 &      0.17 &      0.13 &      0.21 &      0.10 &      0.17 &      0.05 &      0.21 &      0.06 &      0.09 &      0.16 & $<$  3.13 \\
 &  & (  0.20) & (  0.01) & (  0.00) & (  0.00) & (  0.01) & (  0.02) & (  0.03) & (  0.02) & (  0.01) & (  0.01) & (  0.00) \\
$[$NeV$]$    & 3346 &      0.12 &      0.15 &      0.25 &      0.11 &      0.24 &      0.17 &      0.18 &      0.16 &      0.10 &      0.02 &      0.18 \\
 &  & (  0.10) & (  0.01) & (  0.00) & (  0.01) & (  0.01) & (  0.02) & (  0.03) & (  0.02) & (  0.01) & (  0.01) & (  0.28) \\
$[$NeV$]$    & 3426 &      0.35 &      0.49 &      0.65 &      0.41 &      0.61 &      0.44 &      0.61 &      0.25 &      0.23 &      0.33 &      0.18 \\
 &  & (  0.01) & (  0.01) & (  0.01) & (  0.01) & (  0.02) & (  0.02) & (  0.08) & (  0.04) & (  0.02) & (  0.02) & (  0.00) \\
$[$OII$]$    & 3727 &      2.94 &      1.22 &      1.13 &      1.45 &      1.59 &      1.66 &      0.71 &      1.54 &      1.89 &      1.57 &      1.73 \\
 &  & (  0.09) & (  0.02) & (  0.01) & (  0.03) & (  0.05) & (  0.04) & (  0.19) & (  0.18) & (  0.12) & (  0.08) & (  0.03) \\
$[$NeIII$]$  & 3869 &      1.03 &      1.06 &      1.07 &      0.86 &      0.95 &      0.94 &      0.97 &      0.67 &      0.86 &      0.82 &      0.80 \\
 &  & (  0.03) & (  0.05) & (  0.01) & (  0.02) & (  0.03) & (  0.03) & (  0.14) & (  0.11) & (  0.05) & (  0.04) & (  0.02) \\
H8       & 3889 &      0.11 &      0.12 &      0.12 &      0.10 &      0.24 &      0.07 &      0.14 &      0.09 &      0.10 & $<$  0.02 &      0.10 \\
 &  & (  0.01) & (  0.02) & (  0.00) & (  0.01) & (  0.23) & (  0.03) & (  0.02) & (  0.03) & (  0.01) & (  0.00) & (  0.02) \\
$[$NeIII$]$  & 3968 &      0.46 &      0.41 &      0.40 &      0.32 &      0.37 &      0.27 &      0.38 &      0.22 &      0.36 &      0.25 &      0.35 \\
 &  & (  0.04) & (  0.02) & (  0.01) & (  0.03) & (  0.03) & (  0.03) & (  0.05) & (  0.04) & (  0.03) & (  0.03) & (  0.02) \\
$[$SII$]$    & 4074 &      0.06 &      0.16 &      0.15 &      0.07 &      0.20 &      0.12 &      0.23 &      0.13 &      0.35 &      0.27 &      0.33 \\
 &  & (  0.03) & (  0.01) & (  0.01) & (  0.02) & (  0.02) & (  0.03) & (  0.03) & (  0.03) & (  0.03) & (  0.03) & (  0.02) \\
H$\delta$  & 4102 &      0.15 &      0.15 &      0.17 &      0.13 &      0.14 &      0.17 &      0.16 &      0.11 &      0.12 &      0.13 &      0.15 \\
 &  & (  0.04) & (  0.01) & (  0.01) & (  0.02) & (  0.02) & (  0.02) & (  0.02) & (  0.03) & (  0.02) & (  0.02) & (  0.02) \\
H$\gamma$  & 4340 &      0.30 &      0.35 &      0.35 &      0.29 &      0.36 &      0.34 &      0.37 &      0.29 &      0.39 &      0.32 &      0.34 \\
 &  & (  0.03) & (  0.02) & (  0.01) & (  0.04) & (  0.02) & (  0.03) & (  0.05) & (  0.05) & (  0.03) & (  0.03) & (  0.01) \\
$[$OIII$]$   & 4363 &      0.16 &      0.18 &      0.17 &      0.22 &      0.18 &      0.24 &      0.25 &      0.16 &      0.14 &      0.13 &      0.17 \\
 &  & (  0.03) & (  0.02) & (  0.01) & (  0.03) & (  0.02) & (  0.02) & (  0.03) & (  0.03) & (  0.02) & (  0.01) & (  0.01) \\
HeII     & 4686 &      0.24 &      0.23 &      0.28 &      0.14 &      0.22 &      0.20 &      0.30 &      0.13 &      0.15 &      0.15 &      0.25 \\
 &  & (  0.01) & (  0.00) & (  0.01) & (  0.01) & (  0.01) & (  0.01) & (  0.04) & (  0.01) & (  0.01) & (  0.01) & (  0.05) \\
H$\beta$   & 4861 &      1.00 &      1.00 &      1.00 &      1.00 &      1.00 &      1.00 &      1.00 &      1.00 &      1.00 &      1.00 &      1.00 \\
 &  & (  0.03) & (  0.02) & (  0.01) & (  0.02) & (  0.04) & (  0.02) & (  0.17) & (  0.12) & (  0.09) & (  0.05) & (  0.02) \\
$[$OIII$]$   & 4959 &      4.15 &      4.93 &      5.35 &      4.57 &      4.78 &      5.11 &      5.59 &      3.83 &      3.97 &      3.80 &      3.55 \\
 &  & (  0.23) & (  0.63) & (  0.04) & (  0.08) & (  0.16) & (  0.12) & (  0.86) & (  0.56) & (  0.25) & (  0.20) & (  0.10) \\
$[$OIII$]$   & 5007 &     12.73 &     14.95 &     16.37 &     14.02 &     14.90 &     14.55 &     17.05 &     11.63 &     11.77 &     11.97 &     10.37 \\
 &  & (  0.32) & (  0.45) & (  0.12) & (  0.25) & (  0.62) & (  0.51) & (  2.39) & (  1.55) & (  0.73) & (  0.64) & (  0.25) \\
$[$FeVII$]$  & 5159 &      0.14 &      0.03 &      0.04 &      0.12 &      0.04 &      0.09 &      0.03 &      0.08 &      0.09 &      0.08 &      0.11 \\
 &  & (  0.02) & (  0.01) & (  0.01) & (  0.03) & (  0.01) & (  0.02) & (  0.01) & (  0.02) & (  0.01) & (  0.02) & (  0.01) \\
$[$NI$]$     & 5200 &      0.16 &      0.07 &      0.08 &      0.10 &      0.10 &      0.22 &      0.09 &      0.20 &      0.21 &      0.26 &      0.21 \\
 &  & (  0.02) & (  0.01) & (  0.01) & (  0.02) & (  0.01) & (  0.01) & (  0.03) & (  0.02) & (  0.02) & (  0.02) & (  0.01) \\
$[$FeVII$]$  & 5722 &      0.17 &      0.08 &      0.10 & $<$  0.03 &      0.12 &      0.05 &      0.08 &      0.09 &      0.02 & $<$  0.04 &      0.03 \\
 &  & (  0.02) & (  0.01) & (  0.00) & (  0.00) & (  0.01) & (  0.02) & (  0.01) & (  0.01) & (  0.00) & (  0.00) & (  0.00) \\
$[$NII$]$    & 5755 &      0.15 &      0.09 &      0.05 &      0.11 &      0.07 &      0.13 &      0.10 &      0.09 &      0.11 &      0.16 &      0.12 \\
 &  & (  0.02) & (  0.01) & (  0.00) & (  0.01) & (  0.01) & (  0.01) & (  0.02) & (  0.01) & (  0.01) & (  0.01) & (  0.00) \\
HeI      & 5876 &      0.23 &      0.11 &      0.11 &      0.14 &      0.07 &      0.16 &      0.07 &      0.16 &      0.12 &      0.18 &      0.17 \\
 &  & (  0.03) & (  0.01) & (  0.00) & (  0.00) & (  0.01) & (  0.01) & (  0.02) & (  0.03) & (  0.01) & (  0.01) & (  0.00) \\
$[$FeVII$]$  & 6087 & ..... &      0.11 &      0.13 & $<$  0.15 &      0.16 &      0.15 &      0.16 &      0.11 &      0.06 &      0.04 &      0.05 \\
 &  & (.....) & (  0.01) & (  0.01) & (  0.00) & (  0.02) & (  0.02) & (  0.02) & (  0.03) & (  0.01) & (  0.02) & (  0.01) \\
$[$OI$]$     & 6300 &      1.02 &      0.69 &      0.77 &      0.86 &      1.04 &      1.40 &      1.50 &      1.32 &      1.89 &      1.73 &      1.45 \\
 &  & (  0.03) & (  0.02) & (  0.01) & (  0.02) & (  0.03) & (  0.06) & (  0.24) & (  0.16) & (  0.12) & (  0.08) & (  0.03) \\
$[$OI$]$     & 6363 &      0.47 &      0.22 &      0.25 &      0.30 &      0.38 &      0.47 &      0.54 &      0.42 &      0.58 &      0.69 &      0.49 \\
 &  & (  0.02) & (  0.01) & (  0.00) & (  0.02) & (  0.02) & (  0.03) & (  0.09) & (  0.05) & (  0.04) & (  0.04) & (  0.01) \\
$[$NII$]$    & 6548 &      1.98 &      1.27 &      1.17 &      2.25 &      1.62 &      2.59 &      1.94 &      2.21 &      2.64 &      2.66 &      2.10 \\
 &  & (  0.08) & (  0.10) & (  0.01) & (  0.14) & (  0.05) & (  0.04) & (  0.34) & (  0.38) & (  0.20) & (  0.22) & (  0.05) \\
H$\alpha$  & 6563 &      4.84 &      4.37 &      3.29 &      3.74 &      3.75 &      5.61 &      6.08 &      4.25 &      5.11 &      5.10 &      4.31 \\
 &  & (  0.14) & (  0.10) & (  0.26) & (  0.22) & (  0.18) & (  0.21) & (  1.18) & (  0.90) & (  0.55) & (  0.40) & (  0.25) \\
$[$NII$]$    & 6584 &      5.89 &      3.65 &      3.51 &      6.77 &      4.93 &      7.78 &      5.85 &      5.80 &      7.96 &      8.40 &      6.34 \\
 &  & (  0.19) & (  0.13) & (  0.03) & (  0.43) & (  0.16) & (  0.12) & (  0.80) & (  0.86) & (  0.60) & (  0.52) & (  0.16) \\
HeI      & 7065 & $<$  0.05 &      0.05 &      0.08 &      0.05 &      0.15 &      0.18 &      0.13 &      0.10 &      0.10 &      0.04 &      0.05 \\
 &  & (  0.00) & (  0.00) & (  0.01) & (  0.02) & (  0.01) & (  0.01) & (  0.02) & (  0.01) & (  0.01) & (  0.01) & (  0.01) \\
$[$ArIII$]$  & 7135 &      0.23 &      0.36 &      0.36 &      0.33 &      0.42 &      0.46 &      0.56 &      0.26 &      0.62 &      0.32 &      0.41 \\
 &  & (  0.04) & (  0.01) & (  0.01) & (  0.02) & (  0.01) & (  0.01) & (  0.08) & (  0.04) & (  0.04) & (  0.02) & (  0.01) \\
$[$OII$]$    & 7325 &      0.32 &      0.46 &      0.43 &      0.31 &      0.67 &      0.40 &      0.83 &      0.42 &      1.17 &      0.70 &      0.86 \\
 &  & (  0.15) & (  0.02) & (  0.00) & (  0.01) & (  0.02) & (  0.01) & (  0.12) & (  0.06) & (  0.07) & (  0.04) & (  0.06) \\
$[$ArIII$]$  & 7751 &      0.30 &      0.16 &      0.13 &      0.11 &      0.16 &      0.07 &      0.16 &      0.08 &      0.17 &      0.02 &      0.08 \\
 &  & (  0.02) & (  0.02) & (  0.00) & (  0.00) & (  0.01) & (  0.01) & (  0.02) & (  0.01) & (  0.01) & (  0.01) & (  0.01) \\
$[$SIII$]$   & 9069 &      1.42 &      1.21 &      1.26 &      1.12 &      1.57 &      1.79 &      2.12 &      1.10 &      1.83 &      1.37 &      1.25 \\
 &  & (  0.04) & (  0.06) & (  0.01) & (  0.02) & (  0.04) & (  0.03) & (  0.30) & (  0.18) & (  0.11) & (  0.07) & (  0.12) \\
$[$SIII$]$   & 9532 &      2.42 &      2.41 &      2.55 &      2.25 &      4.21 &      3.47 &      5.01 &      2.58 &      5.17 &      2.60 &      3.44 \\
 &  & (  0.08) & (  0.05) & (  0.03) & (  0.07) & (  0.11) & (  0.06) & (  0.68) & (  0.44) & (  0.32) & (  0.14) & (  0.08) \\
\\
f$_{H\beta}$\tablenotemark{f} & 4861 &      1.62 &     10.12 &      6.52 &      2.89 &      3.72 &      2.96 &      4.69 &      4.69 &      6.91 &      3.73 &      7.09 \\
 &  & (  0.04) & (  0.14) & (  0.05) & (  0.05) & (  0.09) & (  0.04) & (  0.56) & (  0.40) & (  0.42) & (  0.14) & (  0.10) \\

\enddata
\tablenotetext{a}{Projected angular distance from nucleus in arcseconds.}
\tablenotetext{b}{Measurement is blue-shifted (b) or red-shifted(r)
 with respect to the Markarian~3 rest-frame.}
\tablenotetext{c}{Angular width of measurement bin in arcseconds.}
\tablenotetext{d}{The 1$\sigma$ measurement uncertainty is listed in parenthesis below
 each measurement}
\tablenotetext{e}{$<$ indicates that the recorded value
is the 1$\sigma$ upper limit.}
\tablenotetext{f}{H$\beta$ fluxes are in units of
  10$^{-15}$ ergs sec$^{-1}$ cm$^{-2}$.}
\end{deluxetable}


\newpage
%
%
\begin{deluxetable}{lrrrrrrrrrrrr}
\tabletypesize{\footnotesize}
\rotate
\tablewidth{530pt}
\tablenum{3}
\tablecolumns{13}
\tablecaption{Extinction Corrected Emission-line Fluxes Relative to H$\beta$\label{tab:3}}
\tablehead{
\colhead{} &
\colhead{} &
\multicolumn{6}{c}{West} &
\multicolumn{5}{c}{East} \\
\colhead{} &
\colhead{} &
\colhead{   -0.8\tablenotemark{a}} &
\colhead{   -0.5} &
\colhead{   -0.2} &
\colhead{   -0.2} &
\colhead{    0.0} &
\colhead{    0.0} &
\colhead{    0.3} &
\colhead{    0.3} &
\colhead{    0.5} &
\colhead{    0.5} &
\colhead{    0.8} \\
\colhead{} &
\colhead{} &
\colhead{r\tablenotemark{b}} &
\colhead{r} &
\colhead{r} &
\colhead{b} &
\colhead{r} &
\colhead{b} &
\colhead{r} &
\colhead{b} &
\colhead{r} &
\colhead{b} &
\colhead{r} \\
\colhead{} &
\colhead{} &
\colhead{    0.3\tablenotemark{c}} &
\colhead{    0.3} &
\colhead{    0.2} &
\colhead{    0.2} &
\colhead{    0.3} &
\colhead{    0.3} &
\colhead{    0.3} &
\colhead{    0.3} &
\colhead{    0.3} &
\colhead{    0.3} &
\colhead{    0.3} \\
\colhead{(1)} &
\colhead{(2)} &
\colhead{(3)} &
\colhead{(4)} &
\colhead{(5)} &
\colhead{(6)} &
\colhead{(7)} &
\colhead{(8)} &
\colhead{(9)} &
\colhead{(10)} &
\colhead{(11)} &
\colhead{(12)} &
\colhead{(13)}
}

\startdata

E(B-V) &    &   0.12 &   0.14 &   0.16 &   0.15 &   0.21 &   0.22 &   0.40 &   0.22 &   0.24 &   0.26 &   0.31 \\
 &    &  (0.02)\tablenotemark{d}  &  (0.01)  &  (0.01)  &  (0.05)  &  (0.02)  &  (0.02)  &  (0.02)  &  (0.05)  &  (0.04)  &  (0.06)  &  (0.08)  \\

Ly$\alpha$ & 1216 &     42.54 &     20.81 &     21.30 &     14.08 &     29.68 &     18.58 &     54.35 &      6.78 &     19.97 &      9.90 &     38.97 \\
 &  & ( 13.18) & (  1.48) & (  2.18) & (  8.35) & (  7.08) & (  5.76) & ( 14.72) & (  4.69) & ( 11.55) & (  7.97) & ( 38.35) \\
NV       & 1240 &      0.86 &      0.98 &      1.05 &      0.49 &      1.29 &      0.69 &      2.46 &      0.40 &      0.22 & $<$  0.46\tablenotemark{e} &      0.56 \\
 &  & (  0.28) & (  0.42) & (  0.11) & (  0.32) & (  0.31) & (  0.22) & (  0.66) & (  0.28) & (  0.13) & (  0.00) & (  0.84) \\
CII      & 1335 &      0.42 &      0.13 &      0.05 &      0.16 & $<$  0.22 & $<$  0.32 & $<$  3.27 &      0.02 &      0.06 &      0.06 &      0.69 \\
 &  & (  0.14) & (  0.09) & (  0.01) & (  0.09) & (  0.00) & (  0.00) & (  0.00) & (  0.03) & (  0.04) & (  0.07) & (  1.17) \\
NIV$]$     & 1486 &      0.31 &      0.37 &      0.25 &      0.59 &      0.15 &      0.41 & $<$  0.59 &      0.20 & $<$  0.15 &      0.15 &    ..... \\
 &  & (  0.35) & (  0.03) & (  0.02) & (  0.29) & (  0.04) & (  0.12) & (  0.00) & (  0.11) & (  0.00) & (  0.10) & (.....) \\
CIV      & 1548 &      3.54 &      3.36 &      4.95 &      1.79 &      4.19 &      2.04 &      4.35 &      0.27 &      1.42 &      0.38 &      1.21 \\
 &  & (  0.89) & (  0.27) & (  0.47) & (  0.86) & (  0.81) & (  0.54) & (  1.00) & (  0.19) & (  0.67) & (  0.31) & (  0.97) \\
CIV      & 1551 &      1.77 &      1.68 &      2.47 &      0.89 &      2.09 &      1.02 &      2.16 &      0.14 &      0.70 &      0.19 &      0.60 \\
 &  & (  0.44) & (  0.14) & (  0.23) & (  0.43) & (  0.40) & (  0.27) & (  0.50) & (  0.10) & (  0.33) & (  0.15) & (  0.48) \\
HeII     & 1640 &      1.80 &      1.73 &      2.09 &      1.07 &      1.65 &      1.51 &      2.32 &      0.97 &      1.19 &      1.17 &      1.94 \\
 &  & (  0.44) & (  0.10) & (  0.17) & (  0.49) & (  0.31) & (  0.36) & (  0.52) & (  0.52) & (  0.55) & (  0.75) & (  1.49) \\
OIII$]$    & 1665 &      0.97 &      0.21 &      0.05 &      0.12 &      0.09 &      0.82 & ..... &      0.11 & $<$  0.05 &      0.67 &      0.45 \\
 &  & (  0.24) & (  0.05) & (  0.03) & (  0.12) & (  0.04) & (  0.20) & (.....) & (  0.07) & (  0.00) & (  0.42) & (  0.35) \\
CIII$]$    & 1909 &      4.74 &      2.09 &      1.85 &      2.45 &      0.58 &      0.24 &      4.99 &      0.71 &      0.77 &      1.65 &      3.43 \\
 &  & (  6.71) & (  0.14) & (  0.13) & (  1.03) & (  0.13) & (  0.05) & (  1.07) & (  0.39) & (  0.32) & (  0.97) & (  2.54) \\
CII$]$     & 2324 &      1.66 &      0.62 &      0.57 &      0.69 &      0.59 &      0.64 &      1.13 &      0.60 &      0.81 &      0.68 &      1.84 \\
 &  & (  2.11) & (  0.03) & (  0.04) & (  0.26) & (  0.09) & (  0.13) & (  0.26) & (  0.27) & (  0.31) & (  0.36) & (  1.17) \\
$[$NeIV$]$   & 2424 &      0.76 &      0.78 &      0.91 &      0.50 &      0.76 &      0.46 &      1.17 &      0.22 &      0.38 &      0.31 &      0.16 \\
 &  & (  0.19) & (  0.08) & (  0.06) & (  0.18) & (  0.11) & (  0.09) & (  0.22) & (  0.09) & (  0.13) & (  0.15) & (  0.14) \\
$[$OII$]$    & 2470 & ..... &      0.26 &      0.16 &      0.09 &      0.07 &      0.21 &      0.47 & $<$  0.11 &      0.22 &      0.17 &      0.40 \\
 &  & (.....) & (  0.01) & (  0.01) & (  0.04) & (  0.02) & (  0.04) & (  0.09) & (  0.00) & (  0.08) & (  0.08) & (  0.23) \\
MgII     & 2796 &      1.27 &      0.60 &      0.32 &      0.71 &      0.70 &      0.50 &      0.47 &      0.26 &      0.64 &      0.49 &      0.69 \\
 &  & (  0.21) & (  0.04) & (  0.03) & (  0.24) & (  0.09) & (  0.09) & (  0.11) & (  0.10) & (  0.20) & (  0.22) & (  0.37) \\
MgII     & 2803 &      0.64 &      0.30 &      0.16 &      0.36 &      0.35 &      0.25 &      0.23 &      0.13 &      0.32 &      0.24 &      0.34 \\
 &  & (  0.10) & (  0.02) & (  0.02) & (  0.12) & (  0.05) & (  0.04) & (  0.05) & (  0.05) & (  0.10) & (  0.11) & (  0.19) \\
$[$OIII$]$   & 3133 &      0.29 &      0.24 &      0.39 &      0.19 &      0.34 &      0.11 &      0.63 &      0.13 &      0.19 &      0.35 & $<$  7.43 \\
 &  & (  0.34) & (  0.01) & (  0.02) & (  0.05) & (  0.04) & (  0.04) & (  0.11) & (  0.05) & (  0.06) & (  0.14) & (  0.00) \\
$[$NeV$]$    & 3346 &      0.19 &      0.24 &      0.41 &      0.18 &      0.43 &      0.31 &      0.44 &      0.30 &      0.19 &      0.03 &      0.38 \\
 &  & (  0.15) & (  0.02) & (  0.02) & (  0.05) & (  0.05) & (  0.06) & (  0.08) & (  0.10) & (  0.06) & (  0.03) & (  0.62) \\
$[$NeV$]$    & 3426 &      0.53 &      0.76 &      1.04 &      0.66 &      1.07 &      0.78 &      1.43 &      0.44 &      0.41 &      0.63 &      0.36 \\
 &  & (  0.08) & (  0.03) & (  0.05) & (  0.18) & (  0.12) & (  0.12) & (  0.24) & (  0.16) & (  0.12) & (  0.24) & (  0.17) \\
$[$OII$]$    & 3727 &      4.05 &      1.72 &      1.64 &      2.08 &      2.45 &      2.60 &      1.39 &      2.44 &      3.03 &      2.58 &      3.05 \\
 &  & (  0.57) & (  0.06) & (  0.08) & (  0.55) & (  0.27) & (  0.36) & (  0.40) & (  0.78) & (  0.80) & (  0.93) & (  1.34) \\
$[$NeIII$]$  & 3869 &      1.36 &      1.43 &      1.47 &      1.17 &      1.38 &      1.38 &      1.72 &      0.99 &      1.29 &      1.26 &      1.30 \\
 &  & (  0.19) & (  0.08) & (  0.07) & (  0.30) & (  0.15) & (  0.19) & (  0.30) & (  0.33) & (  0.33) & (  0.44) & (  0.56) \\
H8       & 3889 &      0.14 &      0.16 &      0.17 &      0.14 &      0.35 &      0.11 &      0.24 &      0.13 &      0.15 & $<$  0.02 &      0.17 \\
 &  & (  0.02) & (  0.03) & (  0.01) & (  0.04) & (  0.34) & (  0.04) & (  0.04) & (  0.05) & (  0.04) & (  0.00) & (  0.08) \\
$[$NeIII$]$  & 3968 &      0.59 &      0.53 &      0.53 &      0.42 &      0.52 &      0.38 &      0.63 &      0.32 &      0.52 &      0.37 &      0.54 \\
 &  & (  0.09) & (  0.03) & (  0.03) & (  0.11) & (  0.06) & (  0.07) & (  0.10) & (  0.11) & (  0.13) & (  0.13) & (  0.23) \\
$[$SII$]$    & 4074 &      0.07 &      0.20 &      0.19 &      0.10 &      0.27 &      0.17 &      0.36 &      0.17 &      0.48 &      0.37 &      0.48 \\
 &  & (  0.04) & (  0.02) & (  0.01) & (  0.03) & (  0.04) & (  0.04) & (  0.06) & (  0.07) & (  0.12) & (  0.13) & (  0.20) \\
H$\delta$  & 4102 &      0.18 &      0.19 &      0.21 &      0.17 &      0.18 &      0.22 &      0.25 &      0.15 &      0.16 &      0.18 &      0.21 \\
 &  & (  0.05) & (  0.02) & (  0.01) & (  0.05) & (  0.03) & (  0.04) & (  0.04) & (  0.06) & (  0.05) & (  0.07) & (  0.09) \\
H$\gamma$  & 4340 &      0.34 &      0.40 &      0.41 &      0.34 &      0.43 &      0.41 &      0.49 &      0.34 &      0.47 &      0.39 &      0.43 \\
 &  & (  0.06) & (  0.03) & (  0.02) & (  0.09) & (  0.04) & (  0.06) & (  0.08) & (  0.11) & (  0.11) & (  0.13) & (  0.17) \\
$[$OIII$]$   & 4363 &      0.18 &      0.20 &      0.20 &      0.26 &      0.21 &      0.29 &      0.32 &      0.20 &      0.17 &      0.15 &      0.21 \\
 &  & (  0.04) & (  0.02) & (  0.02) & (  0.07) & (  0.03) & (  0.04) & (  0.05) & (  0.06) & (  0.04) & (  0.05) & (  0.08) \\
HeII     & 4686 &      0.25 &      0.24 &      0.29 &      0.15 &      0.23 &      0.21 &      0.32 &      0.13 &      0.17 &      0.16 &      0.27 \\
 &  & (  0.03) & (  0.01) & (  0.01) & (  0.04) & (  0.02) & (  0.03) & (  0.05) & (  0.04) & (  0.04) & (  0.05) & (  0.11) \\
H$\beta$   & 4861 &      1.00 &      1.00 &      1.00 &      1.00 &      1.00 &      1.00 &      1.00 &      1.00 &      1.00 &      1.00 &      1.00 \\
 &  & (  0.12) & (  0.03) & (  0.04) & (  0.22) & (  0.09) & (  0.12) & (  0.19) & (  0.28) & (  0.23) & (  0.30) & (  0.37) \\
$[$OIII$]$   & 4959 &      4.04 &      4.80 &      5.20 &      4.44 &      4.63 &      4.94 &      5.33 &      3.71 &      3.84 &      3.66 &      3.41 \\
 &  & (  0.51) & (  0.63) & (  0.20) & (  0.96) & (  0.43) & (  0.57) & (  0.94) & (  1.06) & (  0.84) & (  1.09) & (  1.23) \\
$[$OIII$]$   & 5007 &     12.26 &     14.37 &     15.69 &     13.45 &     14.20 &     13.84 &     15.87 &     11.06 &     11.17 &     11.34 &      9.75 \\
 &  & (  1.41) & (  0.56) & (  0.59) & (  2.90) & (  1.35) & (  1.63) & (  2.61) & (  3.07) & (  2.44) & (  3.37) & (  3.50) \\
$[$FeVII$]$  & 5159 &      0.13 &      0.03 &      0.04 &      0.11 &      0.03 &      0.08 &      0.03 &      0.08 &      0.08 &      0.07 &      0.09 \\
 &  & (  0.02) & (  0.01) & (  0.01) & (  0.04) & (  0.01) & (  0.02) & (  0.01) & (  0.02) & (  0.02) & (  0.02) & (  0.03) \\
$[$NI$]$     & 5200 &      0.15 &      0.07 &      0.07 &      0.10 &      0.09 &      0.19 &      0.08 &      0.17 &      0.19 &      0.23 &      0.18 \\
 &  & (  0.02) & (  0.01) & (  0.01) & (  0.03) & (  0.01) & (  0.02) & (  0.02) & (  0.05) & (  0.04) & (  0.07) & (  0.06) \\
$[$FeVII$]$  & 5722 &      0.14 &      0.06 &      0.08 & $<$  0.02 &      0.10 &      0.04 &      0.05 &      0.07 &      0.02 & $<$  0.03 &      0.02 \\
 &  & (  0.02) & (  0.01) & (  0.00) & (  0.00) & (  0.01) & (  0.01) & (  0.01) & (  0.02) & (  0.00) & (  0.00) & (  0.01) \\
$[$NII$]$    & 5755 &      0.12 &      0.07 &      0.04 &      0.09 &      0.06 &      0.10 &      0.07 &      0.07 &      0.08 &      0.12 &      0.09 \\
 &  & (  0.02) & (  0.01) & (  0.00) & (  0.02) & (  0.01) & (  0.01) & (  0.01) & (  0.02) & (  0.02) & (  0.03) & (  0.03) \\
HeI      & 5876 &      0.19 &      0.09 &      0.09 &      0.11 &      0.06 &      0.12 &      0.05 &      0.12 &      0.09 &      0.13 &      0.12 \\
 &  & (  0.03) & (  0.01) & (  0.00) & (  0.02) & (  0.01) & (  0.01) & (  0.01) & (  0.03) & (  0.02) & (  0.04) & (  0.04) \\
$[$FeVII$]$  & 6087 & ..... &      0.08 &      0.10 & $<$  0.12 &      0.12 &      0.11 &      0.10 &      0.08 &      0.04 &      0.03 &      0.03 \\
 &  & (.....) & (  0.01) & (  0.01) & (  0.00) & (  0.01) & (  0.02) & (  0.02) & (  0.03) & (  0.01) & (  0.01) & (  0.01) \\
$[$OI$]$     & 6300 &      0.77 &      0.52 &      0.57 &      0.63 &      0.72 &      0.96 &      0.88 &      0.90 &      1.28 &      1.15 &      0.92 \\
 &  & (  0.08) & (  0.02) & (  0.02) & (  0.12) & (  0.06) & (  0.10) & (  0.16) & (  0.23) & (  0.25) & (  0.31) & (  0.29) \\
$[$OI$]$     & 6363 &      0.35 &      0.16 &      0.18 &      0.22 &      0.26 &      0.32 &      0.31 &      0.29 &      0.39 &      0.45 &      0.30 \\
 &  & (  0.04) & (  0.01) & (  0.01) & (  0.04) & (  0.02) & (  0.04) & (  0.06) & (  0.07) & (  0.08) & (  0.12) & (  0.10) \\
$[$NII$]$    & 6548 &      1.45 &      0.91 &      0.82 &      1.58 &      1.08 &      1.70 &      1.06 &      1.44 &      1.70 &      1.68 &      1.26 \\
 &  & (  0.15) & (  0.08) & (  0.03) & (  0.32) & (  0.09) & (  0.17) & (  0.20) & (  0.39) & (  0.34) & (  0.45) & (  0.40) \\
H$\alpha$  & 6563 &      3.52 &      3.12 &      2.30 &      2.63 &      2.48 &      3.67 &      3.31 &      2.77 &      3.28 &      3.21 &      2.56 \\
 &  & (  0.36) & (  0.10) & (  0.20) & (  0.52) & (  0.22) & (  0.38) & (  0.69) & (  0.83) & (  0.70) & (  0.86) & (  0.82) \\
$[$NII$]$    & 6584 &      4.27 &      2.59 &      2.44 &      4.75 &      3.25 &      5.07 &      3.16 &      3.76 &      5.09 &      5.26 &      3.75 \\
 &  & (  0.44) & (  0.11) & (  0.08) & (  0.94) & (  0.27) & (  0.50) & (  0.49) & (  0.98) & (  1.01) & (  1.38) & (  1.18) \\
HeI      & 7065 & $<$  0.04 &      0.03 &      0.05 &      0.04 &      0.09 &      0.11 &      0.06 &      0.06 &      0.06 &      0.02 &      0.03 \\
 &  & (  0.00) & (  0.00) & (  0.00) & (  0.02) & (  0.01) & (  0.01) & (  0.01) & (  0.01) & (  0.01) & (  0.01) & (  0.01) \\
$[$ArIII$]$  & 7135 &      0.16 &      0.24 &      0.23 &      0.21 &      0.25 &      0.27 &      0.26 &      0.15 &      0.35 &      0.18 &      0.22 \\
 &  & (  0.03) & (  0.01) & (  0.01) & (  0.04) & (  0.02) & (  0.03) & (  0.04) & (  0.04) & (  0.07) & (  0.04) & (  0.07) \\
$[$OII$]$    & 7325 &      0.21 &      0.29 &      0.27 &      0.20 &      0.39 &      0.23 &      0.37 &      0.24 &      0.66 &      0.38 &      0.43 \\
 &  & (  0.10) & (  0.01) & (  0.01) & (  0.04) & (  0.03) & (  0.02) & (  0.06) & (  0.06) & (  0.12) & (  0.10) & (  0.13) \\
$[$ArIII$]$  & 7751 &      0.19 &      0.10 &      0.08 &      0.07 &      0.09 &      0.04 &      0.06 &      0.04 &      0.09 &      0.01 &      0.04 \\
 &  & (  0.02) & (  0.01) & (  0.00) & (  0.01) & (  0.01) & (  0.01) & (  0.01) & (  0.01) & (  0.02) & (  0.00) & (  0.01) \\
$[$SIII$]$   & 9069 &      0.77 &      0.63 &      0.63 &      0.57 &      0.71 &      0.79 &      0.65 &      0.48 &      0.78 &      0.56 &      0.46 \\
 &  & (  0.07) & (  0.03) & (  0.02) & (  0.10) & (  0.05) & (  0.07) & (  0.10) & (  0.12) & (  0.14) & (  0.13) & (  0.14) \\
$[$SIII$]$   & 9532 &      1.28 &      1.22 &      1.24 &      1.12 &      1.85 &      1.48 &      1.48 &      1.09 &      2.13 &      1.03 &      1.21 \\
 &  & (  0.12) & (  0.04) & (  0.04) & (  0.19) & (  0.13) & (  0.13) & (  0.22) & (  0.28) & (  0.37) & (  0.24) & (  0.34) \\
\\
f$_{H\beta}$\tablenotemark{f} & 4861 &      4.55 &     30.28 &     20.81 &      9.03 &     14.10 &     11.71 &     33.79 &     18.87 &     28.97 &     16.77 &     38.12 \\
 &  & (  0.38) & (  0.69) & (  0.57) & (  1.41) & (  0.94) & (  0.96) & (  4.56) & (  3.67) & (  4.71) & (  3.59) & (  9.85) \\

\enddata
\tablenotetext{a}{Projected angular distance from nucleus in arcseconds.}
\tablenotetext{b}{Measurement is blue-shifted (b) or red-shifted(r)
 with respect to the Markarian~3 rest-frame.}
\tablenotetext{c}{Angular width of measurement bin in arcseconds.}
\tablenotetext{d}{The 1$\sigma$ measurement uncertainty is listed in parenthesis below
 each measurement}
\tablenotetext{e}{$<$ indicates that the recorded value
is the 1$\sigma$ upper limit.}
\tablenotetext{f}{H$\beta$ fluxes are in units of
  10$^{-15}$ ergs sec$^{-1}$ cm$^{-2}$.}
\end{deluxetable}


\end{document}